\documentclass[11pt,english]{article}
\usepackage[T1]{fontenc}
\usepackage[latin9]{inputenc}
\usepackage{geometry}
\usepackage{amssymb}
\usepackage{graphicx}
\usepackage{esint}

\makeatletter
\newcommand{\lyxaddress}[1]{
\par {\raggedright #1
\vspace{1.4em}
\noindent\par}
}

\usepackage{amstext}

\makeatletter\usepackage{hyperref}\makeatother
\newcommand{\bc}{\begin{center}}\newcommand{\ec}{\end{center}}

\makeatother

\makeatother

\usepackage{babel}
\begin{document}

% \title{Axial-vector distribution amplitudes\\ in the instanton vacuum model}
\title{Nonperturbative features of the axial current}

\author{B.~Z.~Kopeliovich, Iv\'an~Schmidt and M.~Siddikov}

\maketitle

\lyxaddress{\begin{center}
Departamento de F\'isica, Universidad T\'ecnica Federico Santa Mar\'ia,\\
y Instituto de Estudios Avanzados en Ciencias
e Ingenier\'ia,\\ 
y Centro Cient\'{i}fico - Tecnol\'ogico de Valpara\'{i}so,\\
 Casilla 110-V, Valpara\'{i}so,
Chile
\end{center}
}

\begin{abstract}
We study the nonperturbative structure of the axial current and evaluate
the two-point light-cone distribution amplitudes (DA) associated with the correlator
$\int d\xi\, e^{-iq\cdot\xi}\langle0|\bar{\psi}(x)\Gamma\psi(y)J_{\mu}^{5}(\xi)|0\rangle$
within the instanton vacuum model in the leading
order in $\mathcal{O}\left(N_{c}\right)$. Due to the built-in chiral
symmetry, four of eight axial DAs are related
to that for pions. Knowledge of these nonperturbative objects is important for 
phenomenological study of high-energy neutrino interactions 
and semileptonic decays of heavy quarks.
We provide a code for evaluation of these DAs and an  interpolation
formula valid in the region $Q^{2}\lesssim1$~GeV$^{2}$.
\end{abstract}

\section{Introduction}

According to the Standard Model, the weak interaction of leptons
with quarks has a $V$-$A$ structure. While the hadronic properties of
the vector current has been well established in electro-magnetic processes, the structure of the axial current is less
known, especially in the soft regime of low $Q^2$. At high energies and large $Q^2$ 
(small Bjorken $x$) one can rely on the well-known perturbative QCD expressions
for the light-cone distribution functions (DA) derived in~\cite{Fiore:2005bp,Fiore:2005yi,Motyka:2008ac}.
The difference between the DAs for the vector and axial currents reveals
in different signs in front of the quark masses and some helicity
components. However, in certain processes (e.g. neutrino-induced
interactions, semileptonic decays of heavy bosons), which proceed at a low scale
$Q^{2}\lesssim1$~GeV$^{2}$, the
structures of the axial and vector currents are essentially different due to the spontaneous breaking of the chiral symmetry. At small $Q^2\lesssim m_{\pi}^{2}$ the axial current is
dominated by the contribution of the pion pole. Presence of pions also affects
the region of moderate $Q^{2}$: axial-vector meson dominance does not
work because of the $\rho\pi$ cut~\cite{Deck:1964hm,marage,kpssSpin},
which gives a larger contribution to the dispersion relation than
the $a_{1}$ pole. 

A popular tool to describe the longitudinal component of the axial current is the phenomenological PCAC relation, which has been
proposed in the pre-QCD era \cite{nambu,goldstone,treiman,goldman}.
It relates the longitudinal part of the interaction amplitude of
the axial current with that of the pion \cite{adler}. However, PCAC
requires some model assumptions for continuation of the result to
nonzero $Q^{2}\sim1$~GeV$^{2}$ and does not provide any information
about the transverse part of the axial current.

In this paper we evaluate the two-point quark distribution amplitudes
(DAs) of the axial current in the framework of the instanton vacuum
model (see~\cite{Schafer:1996wv,Diakonov:1985eg,Diakonov:1995qy}
and references therein). The important advantage of this model is
the built-in dynamically broken chiral symmetry, which allows to reproduce 
the low-energy chiral structure of QCD. Recently
this model was used for evaluation of the hadronic
structure of the vector current~\cite{Dorokhov:2003kf,Dorokhov:2006qm}, pion DAs~\cite{Esaibegian:1989uj,Petrov:1998kg,Anikin:1999cx,Anikin:2000rq,Dorokhov:2002iu,Dorokhov:2000gu},
correlators of vector and axial currents~\cite{Dorokhov:2003kf,Dorokhov:2005pg,Dorokhov:2004vf},
as well as different low-energy constants. The
nonperturbative structure of the axial current is particularly important
for processes which include soft kinematics or large distance between
quarks. As was mentioned above, currently the axial distributions
amplitudes can be accessed in the neutrino-hadron interactions (both
inclusive~\cite{Fiore:2005bp,Fiore:2005yi} and exclusive~\cite{kss}),
semileptonic decays of heavy quarks and other processes with production
of charged dileptons ($l\nu$) with small invariant mass $M_{l\nu}^{2}\approx0$.
Although there is a number of other processes which get contributions
from the axial DAs (e.g. charged current mediated electroproduction),
such processes are suppressed at small-$Q^{2}$ as $\sim Q^{4}/m_{W}^{4}$
compared to photon-mediated analogs, and for large $Q^{2}\sim m_{W}^{2}$
all the nonperturbative effects in the wave functions become negligible.

The paper is organized as follows. In Section~\ref{sec:IVM} we briefly
overview the basic elements of the instanton vacuum model (IVM) used for further evaluations. In Section~\ref{sec:DA} we present 
definitions for the DAs. In Section~\ref{sec:PCAC-check} we discuss
the constraints on the axial DAs imposed by the chiral symmetry and
PCAC relation within the IVM model. The leading twist distribution
amplitudes $\Phi_{||}$ and $\Phi_{\perp}$, which are the main result
of the present paper, are derived in Section~\ref{sub:Paral}.
In Section~\ref{sub:g_a_tr} we discuss the subleading twist DAs. We summarize 
the observations and make conclusions in Section~\ref{sec:Conclusion}.

\section{Instanton vacuum model}

\label{sec:IVM}The central object of the model is the effective action
for the light quarks in the instanton vacuum, which in the leading
order in $N_{c}$ has the form~\cite{Diakonov:1985eg,Diakonov:1995qy,Goeke:2007j}
\begin{eqnarray}
S-S_{PT}=\int d^{4}x\left(\frac{N}{V}\ln\lambda+2\Phi^{2}(x)+\bar{\psi}\left(\hat{p}+\hat{v}+\hat{a}\gamma_{5}-m-c\bar{L}f\otimes\Phi\cdot\Gamma_{m}\otimes fL\right)\psi\right)\label{eq:effact}
\end{eqnarray}
 where $\Gamma_{m}$ is one of the matrices, $\Gamma_{m}=1,i\vec{\tau},\gamma_{5}$,
or $i\vec{\tau}\gamma_{5}$;~$\psi$ and $\Phi$ are the fields of
constituent quarks and mesons respectively; $N/V$ is the density
of the instanton gas; $\hat{v}\equiv v_{\mu}\gamma^{\mu}$ is the
external vector current corresponding to the photon; $L$ is the gauge
factor, 
\begin{eqnarray}
L\left(x,z\right) & = & P\exp\left(i\int\limits _{z}^{x}d\zeta^{\mu}\left(v_{\mu}(\zeta)+a_{\mu}\left(\zeta\right)\gamma_{5}\right)\right),\label{eq:L-factor}\\
\bar{L}(x,z) & = & \gamma_{0}L(x,z)^{\dagger}\gamma_{0}
\end{eqnarray}
 which provides the gauge covariance of the action~\cite{Terning:1991yt,Bowler:1994ir}.
A formal derivation of the effective action~(\ref{eq:effact}), which
starts from a minimal substitution in the QCD lagrangian and reproduces
the effective action with gauge factors under standard assumptions
like the zero-mode approximation for the single-instanton propagator,
diluteness of the instanton vacuum and using the algorithm outlined
in~\cite{Diakonov:1985eg,Diakonov:1986tv}, may be found in~\cite{Goeke:2007j}.
As was discussed in~\cite{Diakonov:1986tv}, violation of the gauge
symmetry (in the absence of the gauge factors) originates from the
zero mode approximation and is parametrically small, $\sim\rho^{2}/R^{2}\sim0.1$.
An alternative approach which does not rely on the zero-mode approximation
was discussed in~\cite{Pobylitsa:1989uq}. However, it includes direct
resummation of the instanton propagators and demands very complicated
calculations, even for the simplest correlators. A direct comparison
of the two approaches is not trivial because in~\cite{Pobylitsa:1989uq}
a formal expansion parameter is $\mathcal{O}\left(\sqrt{N/VN_{c}}\right)$,
not $\mathcal{O}\left(1/N_{c}\right)$ as in~\cite{Diakonov:1986tv};
however in the leading order numerically both approaches end up with
close results~\cite{Pobylitsa:1989uq}.

Since in this paper we perform calculations in the leading-order in
$\mathcal{O}\left(1/N_{c}\right),$ for the sake of simplicity we
do not include in~(\ref{eq:effact}) the terms with tensor couplings,
which correspond to subleading non-planar diagrams, and neglected
the finite width of the instanton size distribution.The function $f(p)$
in~(\ref{eq:effact}) is the Fourier transform of the zero-mode profile
in the single-instanton background, 
\begin{equation}
f\left(p\right)=2z\left(I_{0}(z)K_{1}(z)-I_{1}(z)K_{0}(z)-\frac{1}{z}I_{1}(z)K_{1}(z)\right)_{z=p\bar{\rho}/2},\label{eq:f-Bessel}
\end{equation}
 with the high-energy asymptotic behaviour 
\[
f(p)\sim\frac{6}{\left(p\bar{\rho}\right)^{3}}.
\]
 For the sake of simplicity, sometimes the dipole\cite{Diakonov:1985eg}
and Gaussian~\cite{Golli:1998rf} parameterizations of the formfactor
are used,
\begin{equation}
f(p)=\left\{ \begin{array}{cc}
L^{2}/\left(L^{2}-p^{2}\right) & {\rm {(dipole),}}\\
\exp\left(-p^{2}/L^{2}\right) & {\rm {(Gaussian),}}
\end{array}\right.\label{eq:f_def}
\end{equation}
 with $L=\sqrt{2}/\bar{\rho}\sim850\,$MeV. In Euclidean space, both
functions are close to~(\ref{eq:f-Bessel}) and differ only at asymptotically
high momenta. For most of the vacuum condensates, all three parametrizations~(\ref{eq:f-Bessel}-\ref{eq:f_def})
give close numbers. However, the distribution amplitudes, which are
\emph{nonlocal} quark operators with light-cone separation, are much
more sensitive to the choice of the formfactor compared to the local
vacuum condensates. Varying the formfactor $f(p)$ completely different
results were obtained for the leading-twist pion DA in~\cite{Petrov:1998kg,Anikin:2000rq,Dorokhov:2006xw}.
This happens because in the Taylor expansion of a nonlocal operator
the higher moments are very sensitive to the endpoint behaviour of
DAs, which, as will be discussed below, is controlled by the large-momentum
asymptotics of the formfactor. In a special limiting case $f(p)=1$,
which corresponds to the Nambu-Jona-Lasinio (NJL) model~\cite{Nambu:1961tp,Nambu:1961fr},
it is possible to eliminate the gauge factors $L,\bar{L}$ by gauge
rotation, so this case corresponds to the local interaction of mesons
with quarks.

In the leading order in $N_{c}$, we have the same Feynman rules as
in perturbative theory, but with a momentum-dependent quark mass $\mu(p)$
in the quark propagator
\begin{eqnarray}
S(p) & = & \frac{1}{\hat{p}-\mu(p)+i0}.
\end{eqnarray}
 The running mass of the constituent quark has a form 
\begin{equation}
\mu(p)=m+M\, f^{2}(p),\label{eq:mu}
\end{equation}
 where $m\approx5$~MeV is the current quark mass, $M\approx350$~MeV
is the dynamical mass generated by the interaction with the instanton
vacuum background. Due to presence of instantons the vector current
- quark coupling is also modified, 
\begin{eqnarray}
\hat{v} & \equiv & v_{\mu}\gamma^{\mu}\Rightarrow\hat{V}=\hat{v}+\hat{V}^{nonl},\label{eq:v_vert}\\
\hat{a} & \equiv & a_{\mu}\gamma^{\mu}\Rightarrow\hat{A}=\hat{a}+\hat{A}^{nonl},\label{eq:a_vert}
\end{eqnarray}
 In addition to the vertices present in perturbative QCD, the model
has nonlocal terms with higher-order couplings of currents and mesons.
The exact expressions for the nonlocal terms $\hat{V}^{nonl},\hat{A}^{nonl}$
depend on the choice of the path in~(\ref{eq:L-factor}), and one
can find in the literature different results~\cite{Dorokhov:2006qm,Anikin:2000rq,Dorokhov:2003kf,Goeke:2007j}.
As was discussed in~\cite{Goeke:2007j}, this ambiguity arises due
to the zero mode approximation employed in the derivation of~(\ref{eq:effact}).
Since in the absence of the gauge links the violation of the chiral
symmetry is parametrically suppressed as $\sim\rho^{2}/R^{2}$, the
ambiguity in the choice of the path will affect the results only within
the same limits. The longitudinal axial DAs are insensitive to this
choice of the path at all. In what follows we employ the parameterizations,
\begin{eqnarray}
\hat{V}_{nonl} & = & v_{\mu}\left(iM\frac{p_{1}^{\mu}+p_{2}^{\mu}}{p_{1}^{2}-p_{2}^{2}}\left(f\left(p_{1}\right)^{2}-f\left(p_{2}\right)^{2}\right)\right),\label{eq:v_nonl}\\
\hat{A}_{nonl} & = & a_{\mu}\left(iM\frac{p_{1}^{\mu}+p_{2}^{\mu}}{p_{1}^{2}-p_{2}^{2}}\left(f\left(p_{1}\right)-f\left(p_{2}\right)\right)^{2}\right),\label{eq:a_nonl}
\end{eqnarray}
 where $p_{1},\, p_{2}$ are the momenta of the initial and final
quarks.

\section{Distribution amplitudes for the axial current}

\label{sec:DA}
\begin{figure}[h]
\begin{centering}
\includegraphics[scale=0.55]{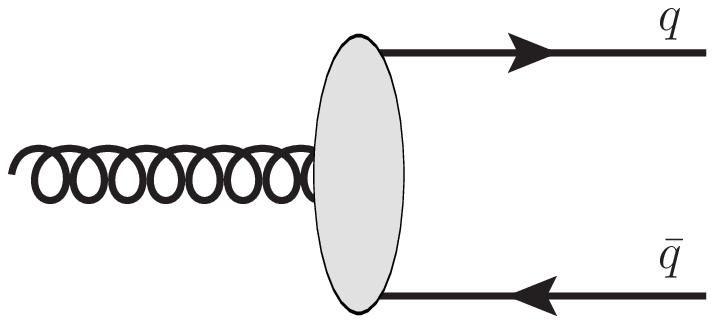}\includegraphics[scale=0.55]{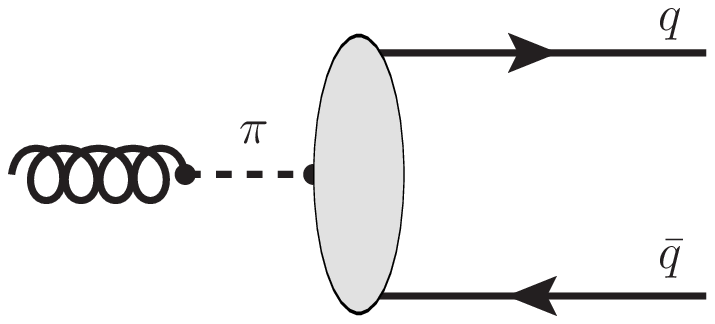} 
\par\end{centering}

\caption{\label{fig:DA_12}The DA contains two terms corresponding to either
intermediate heavy axial states (left), or to a pion (right), which
we label by \textit{bulk} or \textit{pion} respectively.}
\end{figure}

The DAs of the axial current are defined via 3-point
correlators, 
\begin{equation}
\Psi_{\mu}\sim\int d^{4}\xi\, e^{-iq\cdot\xi}\left\langle 0\left|\bar{\psi}\left(y\right)\Gamma\psi\left(x\right)J_{\mu}^{5}(\xi)\right|0\right\rangle ,\label{eq:Psi_mu}
\end{equation}
 where $x$ and $y$ are light-cone coordinates of the quark and antiquark,
$q$ is the momentum flowing through the axial current and $\Gamma$
is one of the Dirac matrices, as was defined in (\ref{eq:effact}).
The structure of the axial current is different from the vector one
because of the spontaneous chiral symmetry breaking and existence
of near-massless pions. In particular, the axial current can fluctuate
into a pion prior to the production of a $\bar{q}q$ pair. Therefore,
the correlator~(\ref{eq:Psi_mu}) consists of two terms, schematically
presented in the Figure~\ref{fig:DA_12}.

One term comes from the combined contribution of the intermediate
heavy states ($a_{1}$ meson, $3\pi$, etc.), and the other one represents
fluctuations of the axial current into a pion. The chiral symmetry
embedded into the model relates the two terms as%
\footnote{A formal proof of this statement is the same as the proof of transversity
of the axial correlator $\langle a_{\mu}a_{\nu}\rangle$ in~\cite{Goeke:2007j}%
}, 
\begin{equation}
\Psi_{\mu}=\Psi_{\mu}^{(bulk)}+\Psi_{\mu}^{(pion)}=\left(g_{\mu\nu}-\frac{q_{\mu}q_{\nu}}{q^{2}-m_{\pi}^{2}}\right)\Psi_{\nu}^{(bulk)}.\label{eq:Psi_nonsing}
\end{equation}
 This form of the DA explicitly satisfies PCAC. In what follows we
concentrate on the part of the amplitude presented in the dispersion
relation for the amplitude by the bulk of heavy states excluding the
pion pole (left pane of Figure~\ref{fig:DA_12}) \cite{kss,kpss},
tacitly assuming that the full DAs are given by Eq.~(\ref{eq:Psi_nonsing}).
We define these DAs as %
\footnote{Sometimes in the literature (see e.g. \cite{Yang:2007zt}) DAs $g_{\perp}^{(v)}$,$h_{||}^{(p)}$~are
defined with prefactor $\epsilon_{\mu\beta\rho\sigma}p_{\rho}z_{\sigma}$
in (\ref{eq:AWF-mu}) and $z_{\beta}$ in (\ref{eq:AWF-5}). This
redefinition corresponds to the change of the DA with its derivative.
We've chosen definitions (\ref{eq:AWF-mu},\ref{eq:AWF-5}) in order
to have relations between functions and not their derivatives in the
PCAC relations (\ref{eq:PCAC-1}-\ref{eq:PCAC-4}).%
}, \\
 
\begin{eqnarray}
\int d^{4}x\, e^{-iqx}\left\langle 0\left|\bar{\psi}\left(-\frac{z}{2}\right)\left[-\frac{z}{2},\frac{z}{2}\right]\gamma_{\mu}\gamma_{5}\psi\left(\frac{z}{2}\right)J_{\beta}^{5}(x)\right|0\right\rangle  & = & if_{A}\int_{0}^{1}d\alpha\, e^{i(0.5-\alpha)p\cdot z}\times\label{eq:AWF-mu5}\\
 & \times & \left(\frac{p_{\mu}z_{\beta}}{p\cdot z}\Phi_{||}(\alpha)+g_{\mu\beta}^{\perp}g_{\perp}^{(a)}(\alpha)\right.\nonumber \\
 & + & \left.\frac{z_{\mu}z_{\beta}}{(p\cdot z)^{2}}g_{3}(\alpha)\right),\nonumber 
\end{eqnarray}

\begin{eqnarray}
\int d^{4}x\, e^{-iqx}\left\langle 0\left|\bar{\psi}\left(-\frac{z}{2}\right)\left[-\frac{z}{2},\frac{z}{2}\right]\gamma_{\mu}\psi\left(\frac{z}{2}\right)J_{\beta}^{5}(x)\right|0\right\rangle  & = & -if_{A}\epsilon_{\mu\beta\rho\sigma}p_{\rho}n_{\sigma}\int_{0}^{1}d\alpha\, e^{i(0.5-\alpha)p\cdot z}\frac{g_{\perp}^{(v)}(\alpha)}{4}\label{eq:AWF-mu}
\end{eqnarray}

\begin{eqnarray}
\int d^{4}x\, e^{-iqx}\left\langle 0\left|\bar{\psi}\left(-\frac{z}{2}\right)\left[-\frac{z}{2},\frac{z}{2}\right]\sigma_{\mu\nu}\gamma_{5}\psi\left(\frac{z}{2}\right)J_{\beta}^{5}(x)\right|0\right\rangle  & = & f_{A}\int_{0}^{1}d\alpha\, e^{i(0.5-\alpha)p\cdot z}\nonumber \\
 & \times & \left(\left(g_{\beta\mu}^{\perp}p_{\nu}-g_{\beta\nu}^{\perp}p_{\mu}\right)\Phi_{\perp}(\alpha)\right.\nonumber \\
 & + & \frac{z_{\beta}}{(p\cdot z)^{2}}\left(p_{\mu}z_{\nu}-p_{\nu}z_{\mu}\right)h_{||}^{(t)}(\alpha)\label{eq:AWF-munu}\\
 & + & \left.\frac{1}{2}\left(g_{\beta\mu}^{\perp}z_{\nu}-g_{\beta\nu}^{\perp}z_{\mu}\right)\frac{1}{p\cdot z}h_{3}(\alpha)\right),\nonumber 
\end{eqnarray}

\begin{eqnarray}
\int d^{4}x\, e^{-iqx}\left\langle 0\left|\bar{\psi}\left(-\frac{z}{2}\right)\left[-\frac{z}{2},\frac{z}{2}\right]\gamma_{5}\psi\left(\frac{z}{2}\right)J_{\beta}^{5}(x)\right|0\right\rangle  & = & f_{A}n_{\beta}\int_{0}^{1}d\alpha\, e^{i(0.5-\alpha)p\cdot z}\frac{h_{||}^{(p)}(\alpha)}{2},\label{eq:AWF-5}
\end{eqnarray}
 where $q$ is the 4-momentum carried by the axial current; $\alpha$
and $\bar{\alpha}\equiv1-\alpha$ are the fractional light-cone momenta
carried by the quark and antiquark; $z$ is the light-cone separation
of the quark-antiquark, $z^{2}=0$; $p_{\mu}$ is the ``positive
direction'' vector on the light-cone, $n_{\mu}$ is the ``negative
direction'' vector on the light-cone, with normalization conditions
$p\cdot n=1,\, p^{2}=n^{2}=0$. Transverse dimensions are chosen in
such a way that the vector $q$ has only components in the $p,n$-
plane. Without any loss of generality, in what follows we choose a
system where $q_{+}=q\cdot n=1$. All the DAs~(\ref{eq:AWF-mu5}-\ref{eq:AWF-5})
contain a color gauge link 
\[
\left[-\frac{z}{2},\frac{z}{2}\right]\equiv P\exp\left(i\int_{-z/2}^{z/2}d\zeta_{\mu}A_{\mu}(\zeta)\right).
\]
 In the instanton vacuum the gluon field $A_{\mu}(\zeta)$ is a sum
of the fields of individual instantons and antiinstantons. As was
discussed in~\cite{Polyakov:1997ea}, the contribution of this link
to the twist-two operators is parametrically suppressed as $(\rho/R)^{4}$
and numerically is negligible. However for higher twists this gauge
link contribution might be important. In general, such calculation
is quite complicated and is doable only in the so-called single-instanton
approximation~\cite{Diakonov:1995qy}. In what follows, for the sake
of simplicity we drop the gauge links. For this reason, the higher-twist
DAs should be considered only as an order of magnitude estimates.
The normalization constant $f_{A}$ is a dimensional parameter introduced
in order to make the distribution amplitudes dimensionless. Its value
is fixed by the condition
\begin{equation}
\int_{0}^{1}d\alpha\,\Phi_{||}\left(\alpha,Q^{2}=0\right)=1.\label{eq:fANorm}
\end{equation}

If we define an ``effective'' axial meson state vector $\left|A^{(\lambda)}(q)\right\rangle $
as
\begin{equation}
\left|A^{(\lambda)}(q)\right\rangle =\int d^{4}x\, e^{-iq\cdot x}e_{\beta}^{(\lambda)}(q)J_{\beta}^{5}(x)\left|0\right\rangle ,\label{eq:A_vec}
\end{equation}
 where the polarization vectors $e^{(\lambda)}$ are defined~ as~\cite{Ball:1998sk}
\begin{equation}
e_{\mu}^{(\lambda=||)}=e^{(\lambda)}\cdot n\left(p_{\mu}-\frac{q^{2}}{2}n_{\mu}\right),\label{eq:e_def}
\end{equation}
 
\begin{eqnarray}
p\cdot e_{\mu}^{(\lambda=\perp)} & = & n\cdot e_{\mu}^{(\lambda=\perp)}=0,\label{eq:-1}\\
\left|e^{(\perp)}(q)\right|^{2} & = & -1,\label{eq:}
\end{eqnarray}
 then we may rewrite Eqns.~(\ref{eq:AWF-mu5-alt}-\ref{eq:AWF-5-alt})
in a standard form as distribution amplitudes of the effective axial
meson state~\cite{Yang:2007zt},

\begin{eqnarray}
\left\langle 0\left|\bar{\psi}\left(y\right)\gamma_{\mu}\gamma_{5}\psi\left(x\right)\right|A(q)\right\rangle  & = & if_{A}\int_{0}^{1}d\alpha\, e^{i(0.5-\alpha)p\cdot z}\times\nonumber \\
 & \times & \left(p_{\mu}\frac{e^{(\lambda)}\cdot z}{p\cdot z}\Phi_{||}(\alpha)+e_{\mu}^{(\lambda=\perp)}g_{\perp}^{(a)}(\alpha)+z_{\mu}\frac{e^{(\lambda)}\cdot z}{(p\cdot z)^{2}}g_{3}(\alpha)\right),\label{eq:AWF-mu5-alt}
\end{eqnarray}

\begin{eqnarray}
\left\langle 0\left|\bar{\psi}\left(y\right)\gamma_{\mu}\psi\left(x\right)\right|A(q)\right\rangle  & = & -if_{A}\epsilon_{\mu\nu\rho\sigma}e_{\nu}^{(\lambda)}p_{\rho}z_{\sigma}\int_{0}^{1}d\alpha\, e^{i(0.5-\alpha)p\cdot z}\frac{g_{\perp}^{(v)}(\alpha)}{4}\label{eq:AWF-mu-alt}
\end{eqnarray}

\begin{eqnarray}
\left\langle 0\left|\bar{\psi}\left(y\right)\sigma_{\mu\nu}\gamma_{5}\psi\left(x\right)\right|A(q)\right\rangle  & = & f_{A}\int_{0}^{1}d\alpha\, e^{i(0.5-\alpha)p\cdot z}\left(\left(e_{\mu}^{(\lambda=\perp)}p_{\nu}-e_{\nu}^{(\lambda=\perp)}p_{\mu}\right)\Phi_{\perp}(\alpha)\right.+\nonumber \\
 & + & \frac{e^{(\lambda)}\cdot z}{(p\cdot z)^{2}}\left(p_{\mu}z_{\nu}-p_{\nu}z_{\mu}\right)h_{||}^{(t)}(\alpha)\label{eq:AWF-munu-alt}\\
 & + & \left.\frac{1}{2}\left(e_{\mu}^{(\lambda)}z_{\nu}-e_{\nu}^{(\lambda)}z_{\mu}\right)\frac{1}{p\cdot z}h_{3}(\alpha)\right),\nonumber 
\end{eqnarray}

\begin{eqnarray}
\left\langle 0\left|\bar{\psi}\left(y\right)\gamma_{5}\psi\left(x\right)\right|A(q)\right\rangle  & = & f_{A}e^{(\lambda)}\cdot n\int_{0}^{1}d\alpha\, e^{i(0.5-\alpha)p\cdot z}\frac{h_{||}^{(p)}(\alpha)}{2}.\label{eq:AWF-5-alt}
\end{eqnarray}

The distribution amplitudes $\Phi_{||}(\alpha),\Phi_{\perp}(\alpha)$
are of twist-2; $g_{\perp}^{(a)},g_{\perp}^{(v)},h_{||}^{(t)},h_{||}^{(p)}$
are of twist-3; $g_{3},h_{3}$ are of twist-4. Chiral parity: all
wave functions in~(\ref{eq:AWF-mu5-alt}), (\ref{eq:AWF-mu-alt})
are chiral even, all wave functions in~(\ref{eq:AWF-munu-alt}),
(\ref{eq:AWF-5-alt}) are chiral odd.

Taking the limit $z\to0$ in (\ref{eq:AWF-mu5}-\ref{eq:AWF-5}) and
using the low-energy expansions dictated for the correlators by chiral
symmetry~\cite{Gasser:1983yg}
\begin{eqnarray*}
i\int d^{4}x\, e^{iqx}\left\langle 0\left|J_{\mu}^{5,a}(x)J_{\nu}^{5,b}(0)\right|0\right\rangle  & = & \delta^{ab}f_{\pi}^{2}\left(g_{\mu\nu}-\frac{q_{\mu}q_{\nu}}{q^{2}-m_{\pi}^{2}}\right)+\mathcal{O}\left(q^{2},m_{\pi}^{2}\right),\\
i\int d^{4}x\, e^{iqx}\left\langle 0\left|J_{\mu}^{5,a}(x)J^{5,b}(0)\right|0\right\rangle  & = & -i\delta^{ab}f_{\pi}G_{\pi}\frac{q_{\mu}}{q^{2}-m_{\pi}^{2}}+\mathcal{O}\left(q^{2},m_{\pi}^{2}\right),\\
G_{\pi}\delta^{ab} & \equiv & \left\langle 0\left|J^{5,a}\right|\pi^{b}(q)\right\rangle =\underbrace{\frac{f_{\pi}m_{\pi}^{2}}{m}}_{2B\, f_{\pi}}\delta^{ab}+\mathcal{O}\left(q^{2},m_{\pi}^{2}\right),
\end{eqnarray*}
 we may obtain the following normalization conditions for the distribution
amplitudes at $Q^{2}=0$:
\begin{equation}
\int d\alpha\,\Phi_{||}(\alpha)=1,\quad\int d\alpha\, g_{a}^{(\perp)}(\alpha)=1,\quad\int d\alpha\, g_{3}(\alpha)=0,\quad\int d\alpha\, h_{||}^{(p)}(\alpha)=1.\label{eq:CL_sum_rule}
\end{equation}

As was discussed in~\cite{Yang:2007zt}, similar to the vector channel,
for axial case approximate Wandzura-Wilczek type relations, 
\begin{eqnarray}
g_{\perp}^{(a),WW}\left(\alpha\right) & \approx & \frac{1}{2}\left[\int_{0}^{u}\frac{dv}{\bar{v}}\Phi_{||}(v)+\int_{u}^{1}\frac{dv}{v}\Phi_{||}(v)\right],\label{eq:WW-1}\\
g_{\perp}^{(v),WW}\left(\alpha\right) & \approx & 2\left[-\int_{0}^{u}\frac{dv}{\bar{v}}\Phi_{||}(v)+\int_{u}^{1}\frac{dv}{v}\Phi_{||}(v)\right],\label{eq:WW-2}\\
h_{||}^{(t),WW}\left(\alpha\right) & \approx & (2\alpha-1)\left[-\int_{0}^{u}\frac{dv}{\bar{v}}\Phi_{\perp}(v)+\int_{u}^{1}\frac{dv}{v}\Phi_{\perp}(v)\right],\label{eq:WW-3}\\
h_{||}^{(p),WW}\left(\alpha\right) & \approx & 2\left[\int_{0}^{u}\frac{dv}{\bar{v}}\Phi_{\perp}(v)-\int_{u}^{1}\frac{dv}{v}\Phi_{\perp}(v)\right],\label{eq:WW-4}
\end{eqnarray}
 can be valid under assumptions that twist-three quark-gluon DAs are
zero.

Let us consider briefly the endpoint behaviour of the DAs ($\alpha$
or $\bar{\alpha}\sim\rho^{2}/R^{2}\sim0.1$ ). In general case, in
the leading order over $1/N_{c}$, the DAs have a structure 
\begin{equation}
\Phi\left(\alpha,q^{2}\right)\sim\int\frac{dl^{-}d^{2}l_{\perp}}{(2\pi)^{3}}\left.\frac{\mathcal{F}(l,\, q)}{\left(l^{2}+\mu^{2}(l)\right)\left(\left(l+q\right)^{2}+\mu^{2}(l+q)\right)}-\frac{\mathcal{G}(l,\, q)}{\left(l^{2}+m^{2}\right)\left(\left(l+q\right)^{2}+m^{2}\right)}\right|_{l^{+}=-\alpha q^{+}},\label{eq:Phi_gen}
\end{equation}
 where the functions $\mathcal{F}(l,q),\,\mathcal{G}(l,\, q)$ depend
on the DA in question, and we assume that 
\begin{eqnarray*}
\mathcal{F}(l,\, q) & = & \mathcal{F}_{0}\left(l^{2},\left(l+q\right)^{2},q^{2}\right)+\mathcal{F}_{1}\left(l^{2},\left(l+q\right)^{2},q^{2}\right)l^{-},\\
\mathcal{G}(l,\, q) & = & \mathcal{G}_{0}\left(l^{2},\left(l+q\right)^{2},q^{2}\right)+\mathcal{G}_{1}\left(l^{2},\left(l+q\right)^{2},q^{2}\right)l^{-}.
\end{eqnarray*}
 In the limit $f(p)\to0$ the running quark mass $\mu(p)\to m$, and
$\mathcal{F}\equiv\mathcal{G}$. First we consider the case when the
function $\mathcal{G}(l,\, q)$ is zero, at least near the endpoints.
In order to analyze the endpoint behaviour of the first term in~(\ref{eq:Phi_gen}),
we follow~\cite{Petrov:1998kg} and for simplicity take $\mu(l)\approx\mu(l+q)\approx M=const$
in the denominator. In this case integration over $l^{-}$ is rather
straightforward and yields
\begin{eqnarray}
\Phi\left(\alpha,q^{2}\right) & \sim & \frac{1}{2q^{+}}\int\frac{d^{2}l_{\perp}}{(2\pi)^{2}}\left(\left.\frac{\mathcal{F}_{0}-\frac{l_{\perp}^{2}+M^{2}}{2\alpha}\mathcal{F}_{1}}{\alpha\bar{\alpha}q^{2}-l_{\perp}^{2}-M^{2}}\right|_{l^{+}=-\alpha q^{+},l^{-}=-\frac{l_{\perp}^{2}+M^{2}}{2\alpha q^{+}}}\right.\label{eq:endpoint}\\
 & - & \left.\left.\frac{\mathcal{G}_{0}-\frac{l_{\perp}^{2}+m^{2}}{2\alpha}\mathcal{G}_{1}}{\alpha\bar{\alpha}q^{2}-l_{\perp}^{2}-m^{2}}\right|_{l^{+}=-\alpha q^{+},l^{-}=-\frac{l_{\perp}^{2}+m^{2}}{2\alpha q^{+}}}\right),\\
\left.l^{2}\right|_{l^{+}=-\alpha q^{+},l^{-}=-\frac{l_{\perp}^{2}+M^{2}}{2\alpha q^{+}}} & = & M^{2},\nonumber \\
\left.\left(l+q\right)^{2}\right|_{l^{+}=-\alpha q^{+},l^{-}=-\frac{l_{\perp}^{2}+M^{2}}{2\alpha q^{+}}} & = & \bar{\alpha}q^{2}-\frac{\bar{\alpha}}{\alpha}\left(l_{\perp}^{2}+M^{2}\right).\nonumber 
\end{eqnarray}
 As one can see, the first term in~(\ref{eq:endpoint}) has a branch
cut which starts at $q^{2}=4M^{2}$. This is due to the fact that
in the instanton vacuum model there is no built-in confinement. In
what follows we assume that $q^{2}<4M^{2}$ and will use~(\ref{eq:endpoint})
in order to discuss the endpoint behaviour of the corresponding DAs
in the following sections. As we will see, this behaviour is extremely
sensitive to the parametrization of the formfactor $f(p)$. In pQCD
limit $\mathcal{F}(l,\, q)\sim1$ or $\sim l_{\perp}^{2}$, so the
integral diverges. For the DAs discussed in this paper $\mathcal{F}_{0}(l,q)$
has a form 
\begin{equation}
\mathcal{F}_{0}\sim f^{m}(l)f^{n}(l+q),\quad m,n\ge0,\label{eq:F0_shape}
\end{equation}
 so the corresponding contribution to $\Phi(l,q)$ in the small-$\alpha$
region is
\begin{equation}
\Phi(\alpha,q)\sim\left\{ \begin{array}{cc}
\alpha^{n}, & {\rm {(Dipole)}}\\
\alpha^{3n/2}, & {\rm {(Parametrization(\ref{eq:f-Bessel}))}}\\
\exp\left(\frac{const}{\alpha}\right), & {\rm {(Gaussian)}}
\end{array}\right..\label{eq:Phi_gen_end}
\end{equation}
 The function $\mathcal{F}_{1}(l,q)$ is a constant (does not depend
on $l$) in most of the integrals discussed below, $\mathcal{F}_{1}(l,q)=\mathcal{F}_{1}$,
so taking into account $\mathcal{F}_{1}=\mathcal{G}_{1}$, we see
that the contribution $\sim\mathcal{F}_{1}$ is finite, 
\[
\Phi(\alpha,q)\sim q^{2}\mathcal{F}_{1}\ln\left(\frac{M^{2}-q^{2}\alpha\bar{\alpha}}{m^{2}-q^{2}\alpha\bar{\alpha}}\right)
\]

The second term in~(\ref{eq:endpoint}) may be integrated directly,
yielding in the infrared region~%
\footnote{The coefficient in front of the log in~(\ref{eq:IR_endpoint}) depends
on the exact structure of $\mathcal{G}_{0,1}$ and will be discussed
below%
} 
\begin{equation}
\sim\ln\left(m^{2}-\alpha\bar{\alpha}q^{2}\right),\label{eq:IR_endpoint}
\end{equation}
 which is logarithmically divergent in the chiral limit and exceeds
the finite endpoint contribution given by~(\ref{eq:Phi_gen_end})
for $\alpha<m^{2}/q^{2}$.

\subsection{On PCAC relation}

\label{sec:PCAC-check}

The PCAC hypothesis was proposed in the pre-QCD era \cite{nambu,goldstone,treiman,goldman}
and has been intensively used as a phenomenological tool for describing
the longitudinal part of the axial hadronic current at small virtualities.
In operator form, the PCAC relation is
\begin{equation}
\partial_{\mu}J_{\mu}^{5,a}=f_{\pi}m_{\pi}^{2}\phi^{a},\label{eq:PCAC-def}
\end{equation}
 where $J_{\mu}^{5,a}$ is the axial current, $\phi^{a}$ is the effective
pion field, and $f_{\pi}$ is the pion decay constant. The PCAC relation~(\ref{eq:PCAC-def})
allows to relate four of eight the axial DAs, Eqs.~(\ref{eq:Phi_parallel_0}),
(\ref{eq:h-parallel-t}), (\ref{eq:h-parallel-p}) and (\ref{eq:g3}),
with corresponding pion DAs, % \footnote{In the Eq. we used $h_{||}^{(p)}(u=0.5)=h_{||}^{(p)}(u=-0.5)$ to

\begin{eqnarray}
f_{A}\Phi_{||}\left(\mathbf{\alpha},\: q^{2}=m_{\pi}^{2}\right) & = & f_{\pi}^{2}\sqrt{2}\phi_{2;\pi}(\alpha)\label{eq:PCAC-1}\\
f_{A}g_{3}\left(\alpha,\: q^{2}=m_{\pi}^{2}\right) & = & \frac{f_{\pi}^{2}\sqrt{2}}{2}\psi_{4;\pi}(\alpha)\label{eq:PCAC-2}\\
f_{A}h_{||}^{(t)}\left(\alpha,\: q^{2}=m_{\pi}^{2}\right) & = & -\frac{\sqrt{2}}{3}\frac{f_{\pi}^{2}m_{\pi}^{2}}{m_{u}+m_{d}}\phi_{3;\pi}^{(\sigma)}(\alpha)\label{eq:PCAC-3}\\
f_{A}h_{||}^{(p)}\left(\alpha,\: q^{2}=m_{\pi}^{2}\right) & = & \frac{2\sqrt{2}f_{\pi}^{2}m_{\pi}^{2}}{m_{u}+m_{d}}\phi_{3;\pi}^{(p)}(\alpha)\label{eq:PCAC-4}
\end{eqnarray}
 where the pion DAs are defined as~\cite{Ball:2006wn,kss}, 
\begin{eqnarray}
\left\langle 0\left|\bar{\psi}\left(y\right)\gamma_{\mu}\gamma_{5}\psi\left(x\right)\right|\pi(q)\right\rangle  & = & if_{\pi}\sqrt{2}\int_{0}^{1}d\alpha\, e^{i(\alpha p\cdot y+\bar{\alpha}p\cdot x)}\times\nonumber \\
 & \times & \left(p_{\mu}\phi_{2;\pi}(\alpha)+\frac{1}{2}\frac{z_{\mu}}{(p\cdot z)}\psi_{4;\pi}(\alpha)\right),\label{eq:piWF-mu5}
\end{eqnarray}

\begin{eqnarray}
\left\langle 0\left|\bar{\psi}\left(y\right)\gamma_{5}\psi\left(x\right)\right|\pi(q)\right\rangle  & = & -if_{\pi}\sqrt{2}\frac{m_{\pi}^{2}}{m_{u}+m_{d}}\int_{0}^{1}d\alpha\, e^{i(\alpha p\cdot y+\bar{\alpha}p\cdot x)}\phi_{3;\pi}^{(p)}(\alpha),\label{eq:piWF-5}
\end{eqnarray}

\begin{eqnarray}
\left\langle 0\left|\bar{\psi}\left(y\right)\sigma_{\mu\nu}\gamma_{5}\psi\left(x\right)\right|\pi(q)\right\rangle  & = & -\frac{i}{3}f_{\pi}\sqrt{2}\frac{m_{\pi}^{2}}{m_{u}+m_{d}}\int_{0}^{1}d\alpha\, e^{i(\alpha p\cdot y+\bar{\alpha}p\cdot x)}\times\nonumber \\
 & \times & \frac{1}{p\cdot z}\left(p_{\mu}z_{\nu}-p_{\nu}z_{\mu}\right)\phi_{3;\pi}^{(\sigma)}(\alpha).\label{eq:piWF-munu}
\end{eqnarray}

It worth to mention that the relations~(\ref{eq:PCAC-1}-\ref{eq:PCAC-4})
may be obtained using instead of (\ref{eq:PCAC-def}) model-independent
$\partial_{\mu}J_{\mu}^{5,a}(x)=2imJ^{5,a}(x)$, where $J^{5,a}(x)$
is pseudoscalar-isovector current, and taking afterwards a residue
at the pion pole. The DAs of the pion Eqs.~(\ref{eq:piWF-mu5})-(\ref{eq:piWF-munu})
were discussed in detail in~\cite{Dorokhov:2003kf,Dorokhov:2006qm,Anikin:2000rq}.
The leading order expressions for these DAs in the framework of the
IVM are presented in Appendix~\ref{sec:pionDAs}.

Since both $\Phi_{||}\left(\alpha,\: q^{2}=m_{\pi}^{2}\right)\approx\Phi_{||}\left(\alpha,\:0\right)$
and $\phi_{2;\pi}(\alpha)$ are normalized to unity, we immediately
conclude that 
\begin{equation}
f_{A}=\sqrt{2}f_{\pi}^{2},\label{eq:fAfPi2}
\end{equation}
 which is confirmed numerically.

In the framework of the instanton vacuum model, the relations~(\ref{eq:PCAC-1}-\ref{eq:PCAC-4})
are satisfied \emph{exactly} due to the built-in chiral symmetry and
transverse structure of the corresponding DAs~(\ref{eq:Psi_nonsing}).
However, in the following sections, we will consider separately the
``transverse'' part $\Psi_{\nu}^{(bulk)}$ and the corresponding
pion DA. The purpose of this exercise is to demonstrate that the DAs
indeed have the transverse structure~(\ref{eq:Psi_nonsing}).

\subsection{Leading-twist distribution amplitudes}

\label{sub:Paral}In the leading twist there are two independent distribution
amplitudes, symmetric $\Phi_{||}(\alpha)$ and antisymmetric $\Phi_{\perp}(\alpha)$.
Straightforward evaluation yields for $\Phi_{||}$ 
\begin{eqnarray}
\Phi_{||}\left(\alpha,\, q^{2}\right) & = & \frac{1}{if_{A}}\int d^{4}x\, e^{-iqx}\int\frac{dz}{2\pi}e^{i(\alpha-0.5)p\cdot z}\left\langle 0\left|\bar{\psi}\left(-\frac{z}{2}\right)\gamma_{+}\gamma_{5}\psi\left(\frac{z}{2}\right)J_{-}^{5}(x)\right|0\right\rangle \label{eq:Phi_parallel_0}\\
 & = & \frac{8N_{c}}{f_{A}}\int\frac{dl^{-}d^{2}l_{\perp}}{(2\pi)^{4}}\left[\frac{\mu(l)\mu(l+q)+l_{\perp}^{2}+\alpha\bar{\alpha}q^{2}}{\left(l^{2}+\mu^{2}(l)\right)\left(\left(l+q\right)^{2}+\mu^{2}(l+q)\right)}-\frac{m^{2}+l_{\perp}^{2}+\alpha\bar{\alpha}q^{2}}{\left(l^{2}+m^{2}\right)\left(\left(l+q\right)^{2}+m^{2}\right)}\right.\nonumber \\
 & - & \left.\frac{M\left(f(l+q)-f(l)\right)^{2}\left(2l^{-}+q^{2}\alpha\right)\left(\mu(l)\bar{\alpha}+\mu(l+q)\alpha\right)}{\left((l+q)^{2}-l^{2}\right)\left(l^{2}+\mu^{2}(l)\right)\left(\left(l+q\right)^{2}+\mu^{2}(l+q)\right)}\right]_{l^{+}=-\alpha q^{+}}\nonumber 
\end{eqnarray}

where the constant $f_{A}$ is fixed from the normalization condition~(\ref{eq:fANorm}),
\begin{eqnarray}
f_{A} & = & 8N_{c}\int\frac{d^{4}l}{(2\pi)^{4}}\left[\frac{\mu^{2}(l)+l_{\perp}^{2}}{\left(l^{2}+\mu^{2}(l)\right)^{2}}-\frac{M\left(f'(l)\right)^{2}l^{-}}{l^{2}\left(l^{2}+\mu^{2}(l)\right)^{2}}\right]
\end{eqnarray}
 As was discussed in the Section~\ref{sec:PCAC-check}, due to PCAC
relation $f_{A}=\sqrt{2}f_{\pi}^{2}$, which is confirmed numerically.

\begin{figure}[htb]
 \bc \includegraphics[scale=0.4]{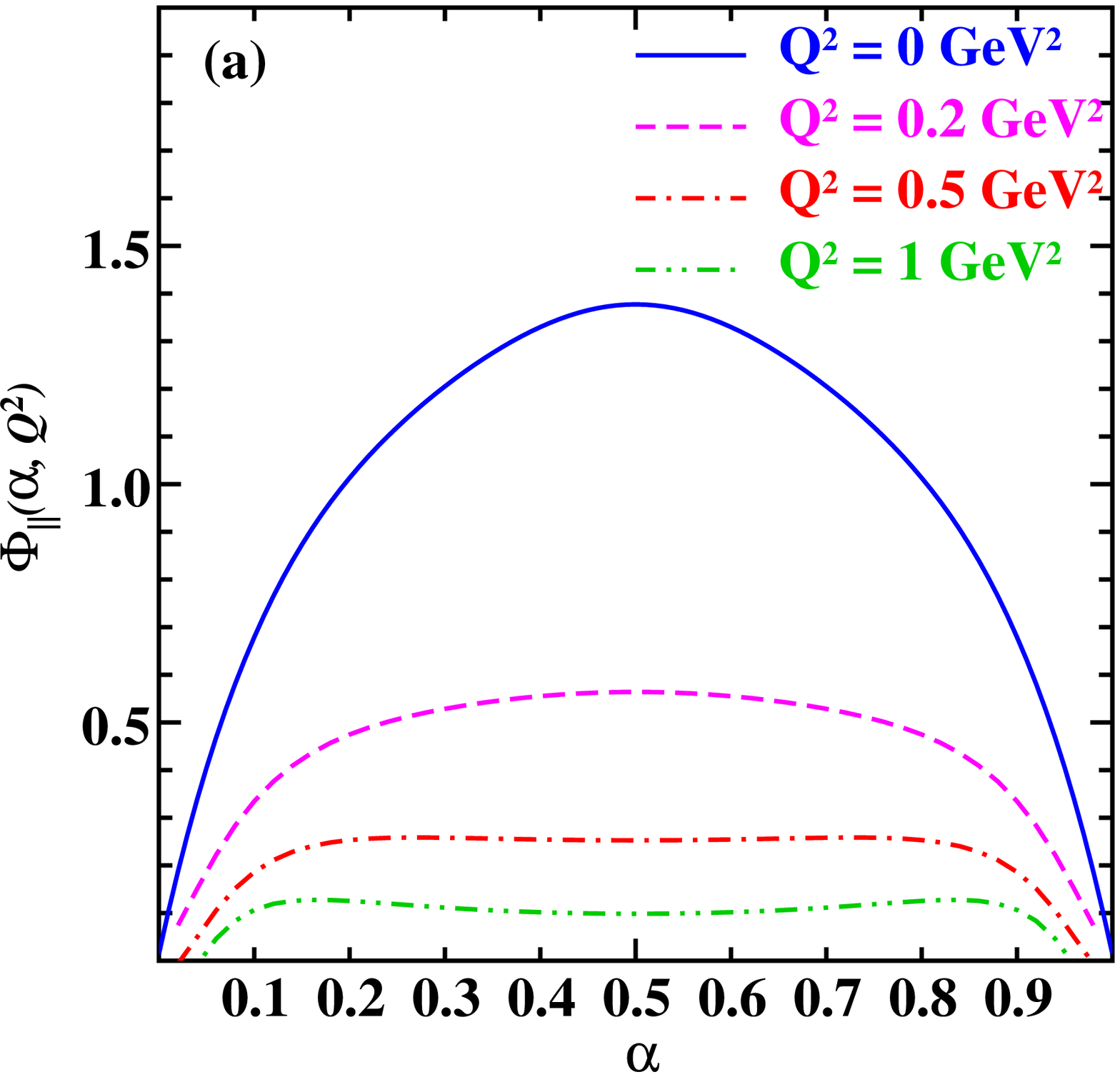} \includegraphics[scale=0.4]{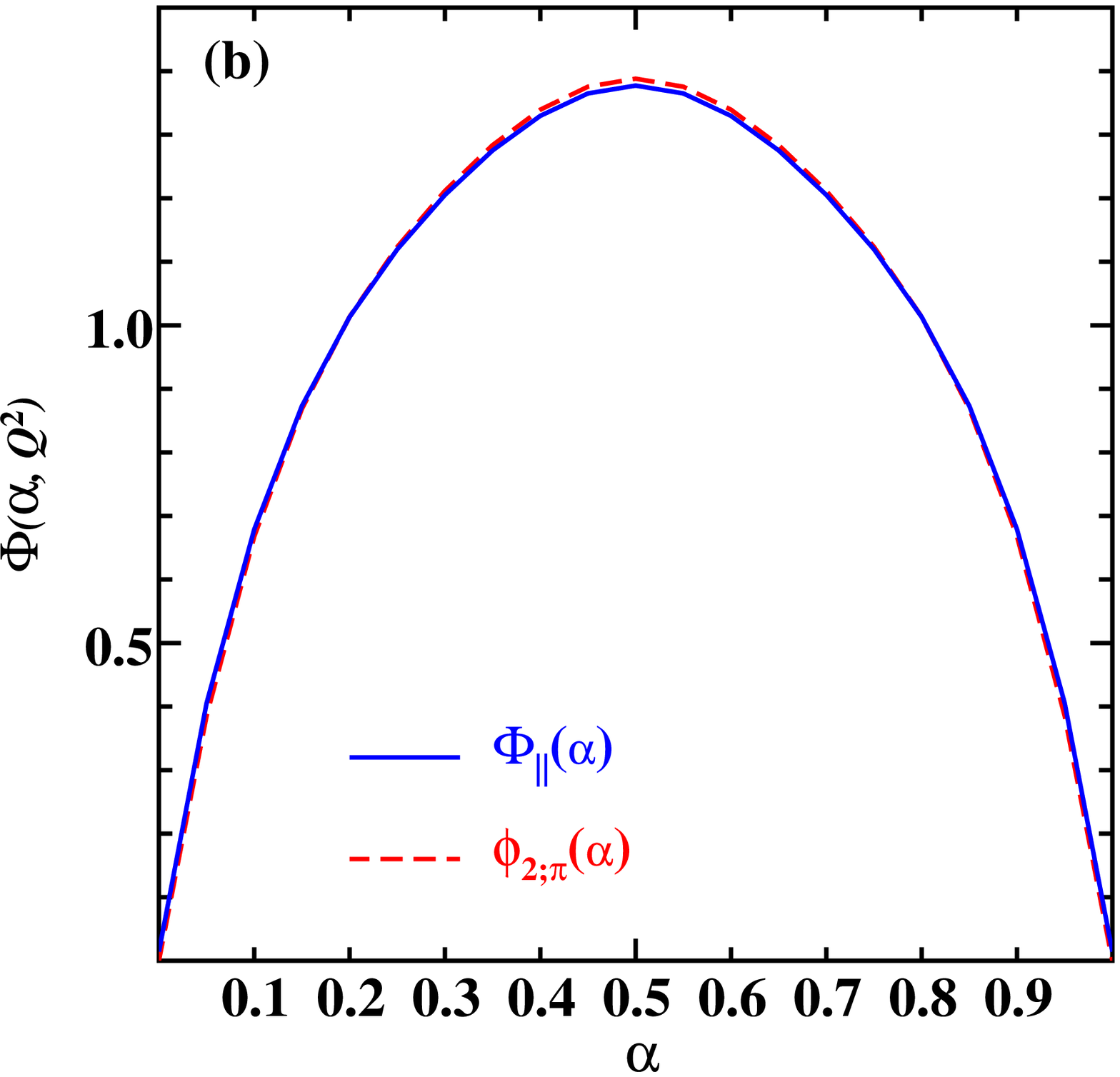}\ec
\caption{\label{fig:Phi_paral}(color online) (a) Dependence of the distribution
amplitude $\Phi_{||}$ on $\alpha$ for several values of $Q^{2}$.
(b) Comparison of the DA $\Phi_{||}$ with pion DA $\phi_{2;\pi}$.
Small deviation from PCAC relation~(\ref{eq:PCAC-1}) is due to finite
numerical precision of the result.}
\end{figure}

In Figure~\ref{fig:Phi_paral} the DA $\Phi_{||}(\alpha)$ is plotted
as function of $\alpha$ for several values of $Q^{2}=-q^{2}$. Near
the endpoints, this distribution amplitude vanishes for all three
parametrizations~(\ref{eq:f-Bessel},\ref{eq:f_def}). Indeed, using
the algorithm discussed at the beginning of this section, the DA may
be approximated with~(\ref{eq:endpoint}), where the function $\mathcal{F}(l,q)$
after some simplifications can be reduced to 
\[
\mathcal{F}(l,\, q)\sim f^{2}(l)f(l+q)\left[f(l+q)-f(l)\right],
\]
 which yields the near-endpoint asymptotics $\Phi_{||}(\alpha)\sim\left(\alpha\bar{\alpha}\right)^{3/2}$
for the parametrization~(\ref{eq:f-Bessel}), $\Phi_{||}(\alpha)\sim\alpha\bar{\alpha}$
for the dipole parametrization, and non-analytic $\Phi_{||}(\alpha)\sim\exp\left(M^{2}/L^{2}\alpha\right)$
for the Gaussian parametrization. As was discussed in Section~\ref{sec:PCAC-check},
this DA is equal to the pion DA $\phi_{2;\pi}(\alpha).$ In the right
pane of the Figure~\ref{fig:Phi_paral} we see that this is indeed
the case (small deviations are due to the finite precision of numerical
calculations). 

\label{sub:Phi_tr}The antisymmetric DA $\Phi_{\perp}(\alpha)$ has
a form

\begin{eqnarray}
\Phi_{\perp}\left(\alpha,\, q^{2}\right) & = & \frac{g_{\beta\mu}n_{\nu}-g_{\beta\nu}n_{\mu}}{4f_{A}}\int d^{4}x\, e^{-iqx}\int\frac{dz}{2\pi}e^{i(\alpha-0.5)p\cdot z}\left\langle 0\left|\bar{\psi}\left(-\frac{z}{2}n\right)\sigma_{\mu\nu}\gamma_{5}\psi\left(\frac{z}{2}n\right)J_{\beta}^{5}(x)\right|0\right\rangle \nonumber \\
 & = & \frac{8N_{c}}{f_{A}}\int\frac{dl_{-}d^{2}l_{\perp}}{(2\pi)^{3}}\left[\frac{-\alpha\mu(l+q)+\bar{\alpha}\mu(l)}{\left(l^{2}+\mu^{2}(l)\right)\left(\left(l+q\right)^{2}+\mu^{2}(l+q)\right)}\right.\label{eq:phi_tr}\\
 & + & \left.\frac{l_{\mu_{\perp}}^{2}}{(l+q)^{2}-l^{2}}\frac{M\left(f(l+q)-f(l)\right)^{2}}{\left(l^{2}+\mu^{2}(l)\right)\left(\left(l+q\right)^{2}+\mu^{2}(l+q)\right)}\right]_{l^{+}=-\alpha q^{+}}\nonumber 
\end{eqnarray}

\begin{figure}[htb]
 \bc \includegraphics[scale=0.4]{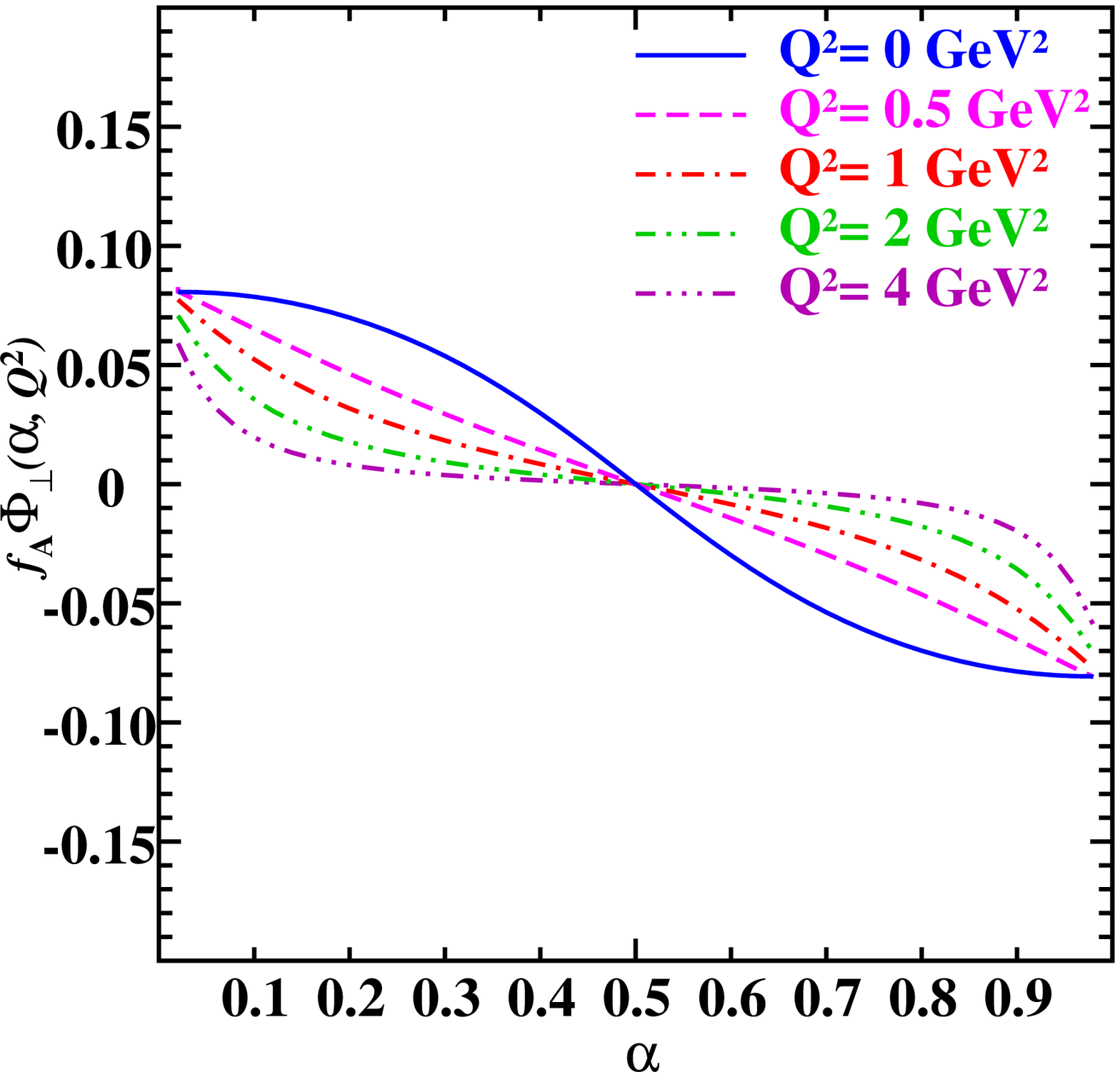} \ec
\caption{\label{fig:Phi_tr}(color online) Dependence of the distribution amplitude
$\Phi_{\perp}$ on $\alpha$ for several values of $Q^{2}$.}
\end{figure}

In Figure~\ref{fig:Phi_tr} the distribution amplitude is shown for
several values of $Q^{2}$. Notice that according to Eq.~(\ref{eq:phi_tr}),
at the endpoints the function does not depend on $q^{2}$ and has
a finite limit.

% In Appendix~\ref{sec:LTDAApprox} we provide numerical expressions
% which may be used for fast estimates of the DAs $\Phi_{||},\,\Phi_{\perp}$.
In Appendix~\ref{sec:LTDAApprox} we provide interpolation formulas for fast estimates of the DAs $\Phi_{||},\,\Phi_{\perp}$.

\subsection{Higher twist distribution amplitudes}

\label{sub:g_a_tr}In twist-three there are four independent distribution
amplitudes $g_{\perp}^{(a)}$, $g_{\perp}^{(v)}$, $h_{||}^{(t)}$,
$h_{||}^{(p)}$. Straightforward evaluation of the twist-3 DAs yields

\begin{eqnarray}
g_{\perp}^{(a)}\left(\alpha,\, q^{2}\right) & = & \frac{8N_{c}}{if_{A}}\int\frac{dl^{-}d^{2}l_{\perp}}{(2\pi)^{4}}\left[\frac{\mu(l)\mu(l+q)+(2\alpha-1)l_{-}q_{+}+\alpha\frac{q^{2}}{2}}{\left(l^{2}+\mu^{2}(l)\right)\left(\left(l+q\right)^{2}+\mu^{2}(l+q)\right)}-\frac{m^{2}+(2\alpha-1)l_{-}q_{+}+\alpha\frac{q^{2}}{2}}{\left(l^{2}+m^{2}\right)\left(\left(l+q\right)^{2}+m^{2}\right)}\right.\label{eq:g_a_tr}\\
 & + & \left.\frac{l_{\mu_{\perp}}^{2}}{(l+q)^{2}-l^{2}}\frac{M\left(f(l+q)-f(l)\right)^{2}(\mu(l+q)-\mu(l))}{\left(l^{2}+\mu^{2}(l)\right)\left(\left(l+q\right)^{2}+\mu^{2}(l+q)\right)}\right]_{l^{+}=-\alpha q^{+}}.\nonumber 
\end{eqnarray}

\begin{eqnarray}
g_{\perp}^{(v)}\left(\alpha,\, q^{2}\right) & = & -\frac{2\epsilon_{\mu\beta\rho\sigma}p_{\rho}n_{\sigma}}{if_{A}}\int d^{4}x\, e^{-iqx}\int\frac{dz}{2\pi}e^{i(\alpha-0.5)p\cdot z}\left\langle 0\left|\bar{\psi}\left(-\frac{z}{2}\right)\gamma_{\mu}\psi\left(\frac{z}{2}\right)J_{\beta}^{5}(x)\right|0\right\rangle \nonumber \\
 & = & \frac{16N_{c}}{f_{A}}\int\frac{dl^{-}d^{2}l_{\perp}}{(2\pi)^{4}}\left[\frac{\alpha\frac{q^{2}}{2}+l_{-}q_{+}}{\left(l^{2}+\mu^{2}(l)\right)\left(\left(l+q\right)^{2}+\mu^{2}(l+q)\right)}-\frac{\alpha\frac{q^{2}}{2}+l_{-}q_{+}}{\left(l^{2}+m^{2}\right)\left(\left(l+q\right)^{2}+m^{2}\right)}\right]_{l^{+}=-\alpha q^{+}}.
\end{eqnarray}

\begin{eqnarray}
h_{||}^{(t)}\left(\alpha,\, q^{2}\right) & = & \frac{\left(p_{\nu}n_{\rho}-p_{\rho}n_{\nu}\right)p_{\beta}}{2f_{A}}\int d^{4}x\, e^{-iqx}\int\frac{dz}{2\pi}e^{i(\alpha-0.5)p\cdot z}\left\langle 0\left|\bar{\psi}\left(-\frac{z}{2}n\right)\sigma_{\nu\rho}\gamma_{5}\psi\left(\frac{z}{2}n\right)J_{\beta}^{5}(x)\right|0\right\rangle \nonumber \\
 & = & -\frac{8N_{c}}{f_{A}}\int\frac{dl^{-}d^{2}l_{\perp}}{(2\pi)^{4}}\left[\frac{\mu(l+q)l_{-}+\mu(l)\left(l^{-}+\frac{q^{2}}{2}\right)}{\left(l^{2}+\mu^{2}(l)\right)\left(\left(l+q\right)^{2}+\mu^{2}(l+q)\right)}-\frac{m\left(2l^{-}+\frac{q^{2}}{2}\right)}{\left(l^{2}+m^{2}\right)\left(\left(l+q\right)^{2}+m^{2}\right)}\right.\label{eq:h-parallel-t}\\
 & + & \left.\frac{2M\left(l^{-}+\frac{q^{2}}{4}\right)\left(l^{-}+\alpha\frac{q^{2}}{2}\right)\left(f(l+q)-f(l)\right)^{2}}{\left((l+q)^{2}-l^{2}\right)\left(l^{2}+\mu^{2}(l)\right)\left(\left(l+q\right)^{2}+\mu^{2}(l+q)\right)}\right]_{l^{+}=-\alpha q^{+}}\nonumber 
\end{eqnarray}

\begin{eqnarray}
h_{||}^{(p)}\left(\alpha,\, q^{2}\right) & = & -\frac{2\: p_{\beta}}{f_{A}}\int d^{4}x\, e^{-iqx}\int\frac{dz}{2\pi}e^{i(\alpha-0.5)p\cdot z}\left\langle 0\left|\bar{\psi}\left(-\frac{z}{2}\right)\gamma_{5}\psi\left(\frac{z}{2}\right)J_{\beta}^{5}(x)\right|0\right\rangle \nonumber \\
 & = & -\frac{16N_{c}}{f_{A}}\int\frac{dl^{-}d^{2}l_{\perp}}{(2\pi)^{4}}\left[\frac{\left(\mu(l)-\mu(l+q)\right)\left(l^{-}+\alpha\frac{q^{2}}{2}\right)}{\left(l^{2}+\mu^{2}(l)\right)\left(\left(l+q\right)^{2}+\mu^{2}(l+q)\right)}\right.\label{eq:h-parallel-p}\\
 & + & \frac{\left(l_{\perp}^{2}+(1-2\alpha)l^{-}-\alpha\frac{q^{2}}{2}+\mu(l)\mu(l+q)\right)}{(l+q)^{2}-l^{2}}\left.\frac{M\left(f(l+q)-f(l)\right)^{2}\left(2l^{-}+q^{2}\alpha\right)}{\left(l^{2}+\mu^{2}(l)\right)\left(\left(l+q\right)^{2}+\mu^{2}(l+q)\right)}\right]_{l^{+}=-\alpha q^{+}}\nonumber 
\end{eqnarray}

\begin{figure}[htb]
 \bc \includegraphics[scale=0.4]{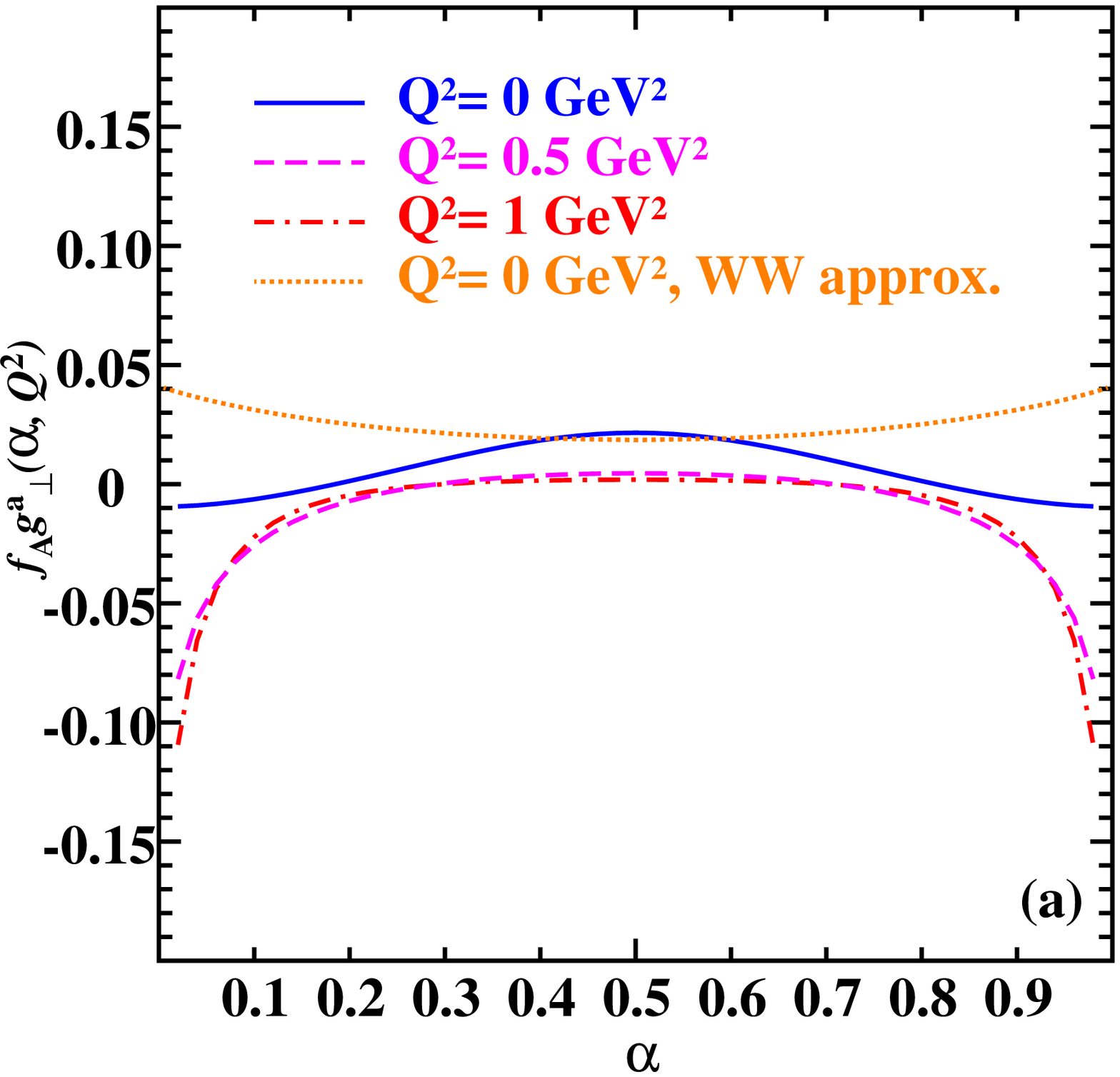}
\includegraphics[scale=0.4]{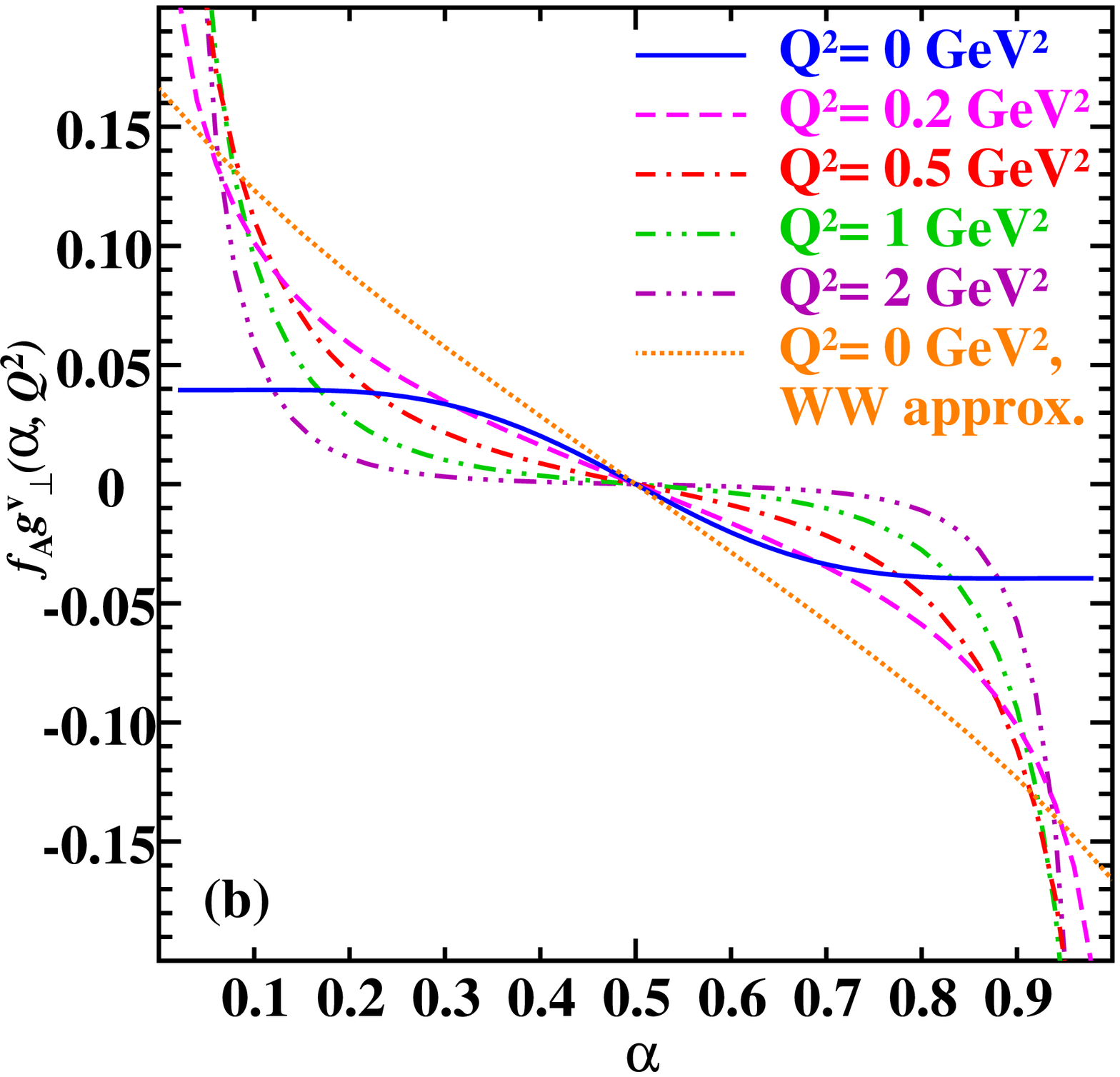} \ec \caption{\label{fig:g_va_tr}(color online) (a) Dependence of the distribution
amplitude $g_{\perp}^{a}$ on $\alpha$ for several values of $Q^{2}$.
(b) Dependence of the distribution amplitude $g_{\perp}^{v}$ on $\alpha$
for several values of $Q^{2}$. In both plots dotted line is a prediction
of the Wandzura-Wilczek approximation~(\ref{eq:WW-1},\ref{eq:WW-2}).}
\end{figure}

\begin{figure}[htb]
\bc \includegraphics[scale=0.4]{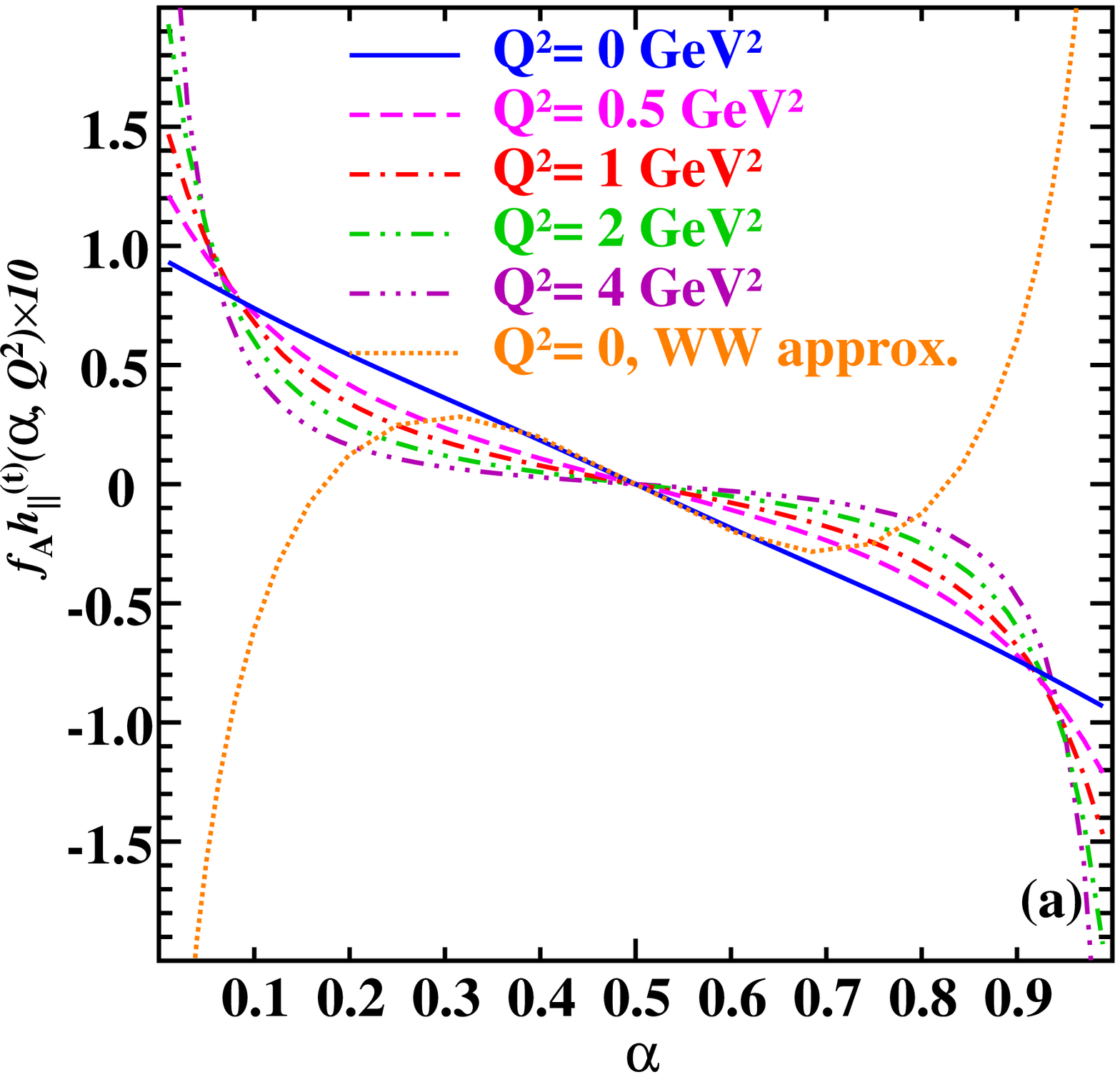} \includegraphics[scale=0.4]{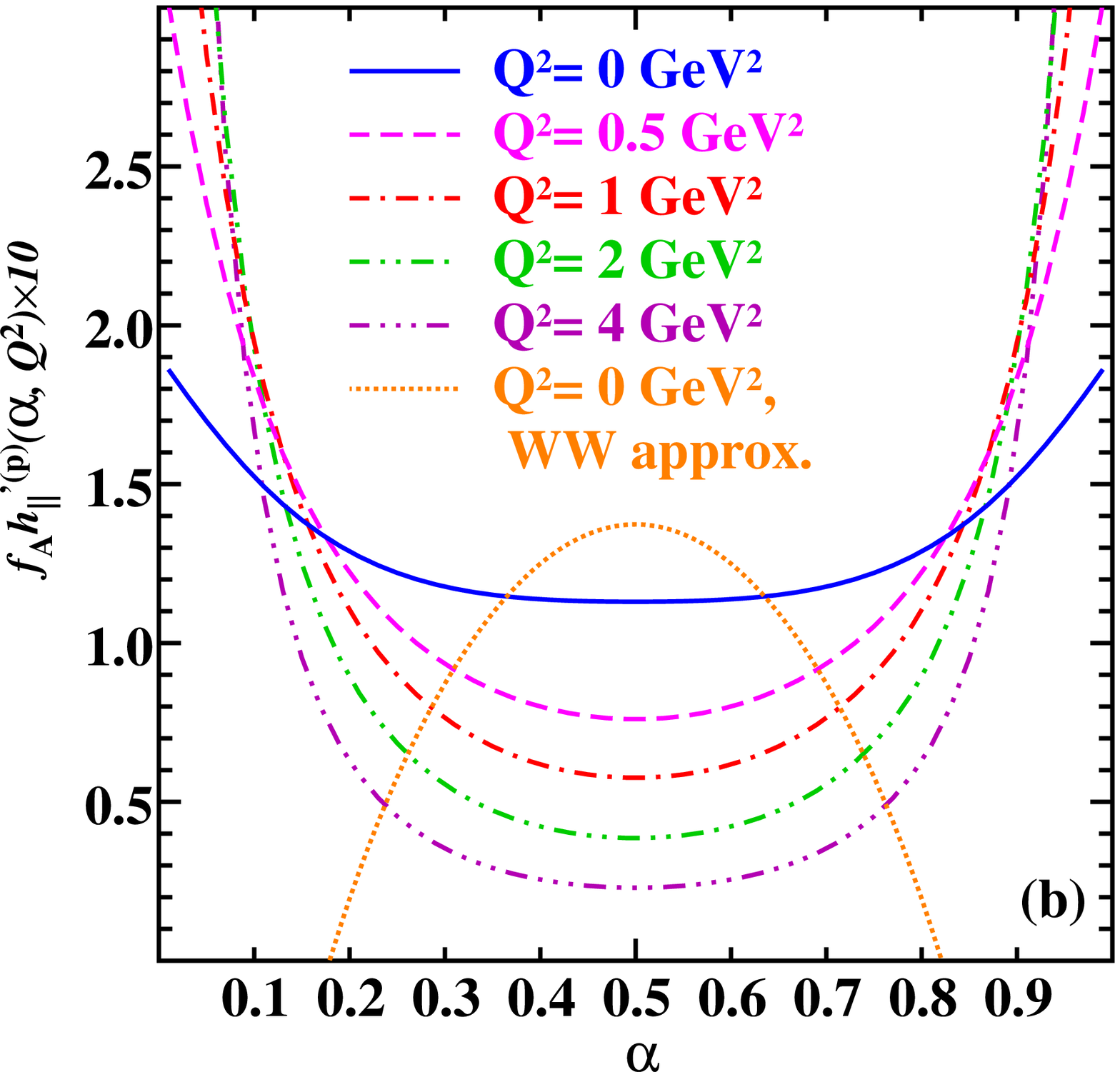}
\ec \caption{\label{fig:h_pt}(color online) (a) Dependence of the distribution
amplitude $h_{||}^{(t)}$ on $\alpha$ for several values of $Q^{2}$.
(b) Dependence of the distribution amplitude $h_{||}^{(p)}$ on $\alpha$
for several values of $Q^{2}$. In both plots dotted line is a prediction
of the Wandzura-Wilczek approximation~(\ref{eq:WW-3},\ref{eq:WW-4}). }
\end{figure}

Using the algorithm discussed at the beginning of this section, we
may get that the endpoint behaviour of these DA is controlled by the
contribution of the type~(\ref{eq:IR_endpoint}) which behaves for
$q^{2}\not=0$ as $\log\left(m^{2}+q^{2}\alpha\bar{\alpha}\right)$,
\begin{equation}
g_{\perp}^{(a)}\left(\alpha,\, q^{2}\right)\sim\left[2m^{2}+q^{2}\left(\alpha^{2}+\bar{\alpha}^{2}\right)\right]\ln\left(m^{2}+q^{2}\alpha\bar{\alpha}\right).\label{eq:g_a_tr_endpoint}
\end{equation}

\begin{equation}
g_{\perp}^{(v)}\left(\alpha,\, q^{2}\right)\sim q^{2}(2\alpha-1)\ln\left(m^{2}+q^{2}\alpha\bar{\alpha}\right).\label{eq:g_v_tr_endpoint}
\end{equation}

\[
h_{||}^{(t)}\left(\alpha,\, q^{2}\right)\sim(2\alpha-1)\log\left(M^{2}+q^{2}\alpha\bar{\alpha}\right).
\]
In the Figures~\ref{fig:g_va_tr},~\ref{fig:h_pt} the distribution
amplitudes~(\ref{eq:g_a_tr}-\ref{eq:h-parallel-p}) are shown as
a function of $\alpha$ for several values of $Q^{2}$. In the same
Figure~\ref{fig:g_va_tr} one can see the prediction of the Wandzura-Wilczek
approximation~(\ref{eq:WW-1}-\ref{eq:WW-4}). As one can see, this
approximation gives results inconsistent with prediction of the present
approach. The reason is that the twist-3 quark-gluon operators $\bar{\psi}(x)\gamma_{\alpha}G_{\mu\nu}(x)[x,y]\psi(y)$
disregarded in the Wandzura-Wilczek approximation are not zero in
the instanton vacuum. Similarly, the Wandzura-Wilczek approximation
fails for the vector current DAs~\cite{Dorokhov:2006qm}. 

While evaluation of the twist-3 DAs was done consistently, for practical
applications the DAs should be evolved to higher scales. In contrast
to a simple ERBL evolution of the leading twist distribution amplitudes,
higher-twist DAs mix with twist-3 three-parton quark-gluon operators.
Although evaluation of the latter is possible in the instanton vacuum~\cite{Diakonov:1995qy},
it is considerably more involved and is beyond the scope of this paper. 

Finally, for the sake of completeness, below we give the expressions
for the twist-4 distribution amplitudes. Up to the best of our knowledge,
there is no any model estimates for these functions. Also, as was
discussed at the beginning of this section, for the twist-4 DAs omission
of the gauge link may be not justified, so these DAs should be considered
as an order-of-magnitude estimates. 

\begin{figure}[htb]
\bc \includegraphics[scale=0.4]{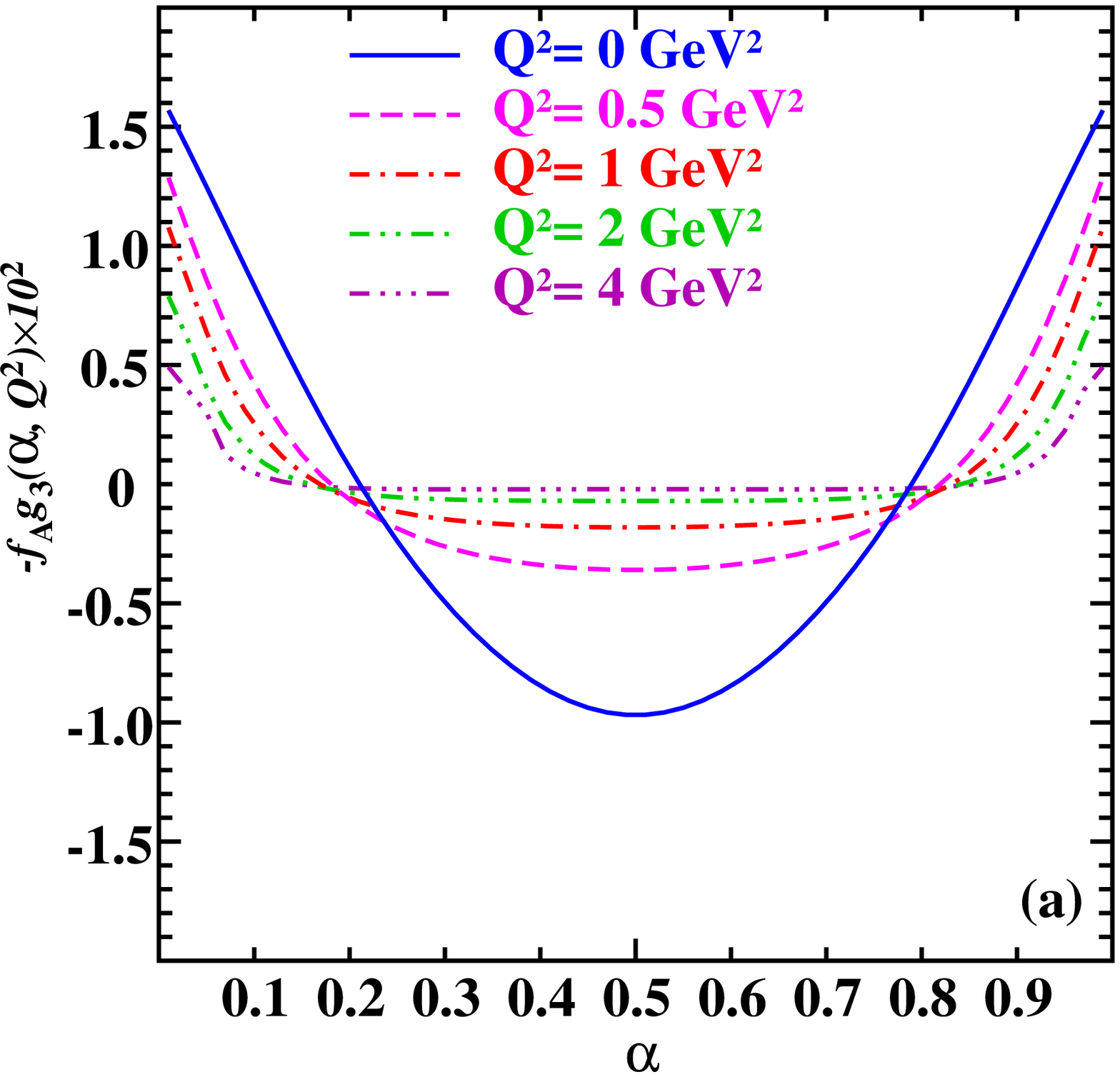} \includegraphics[scale=0.4]{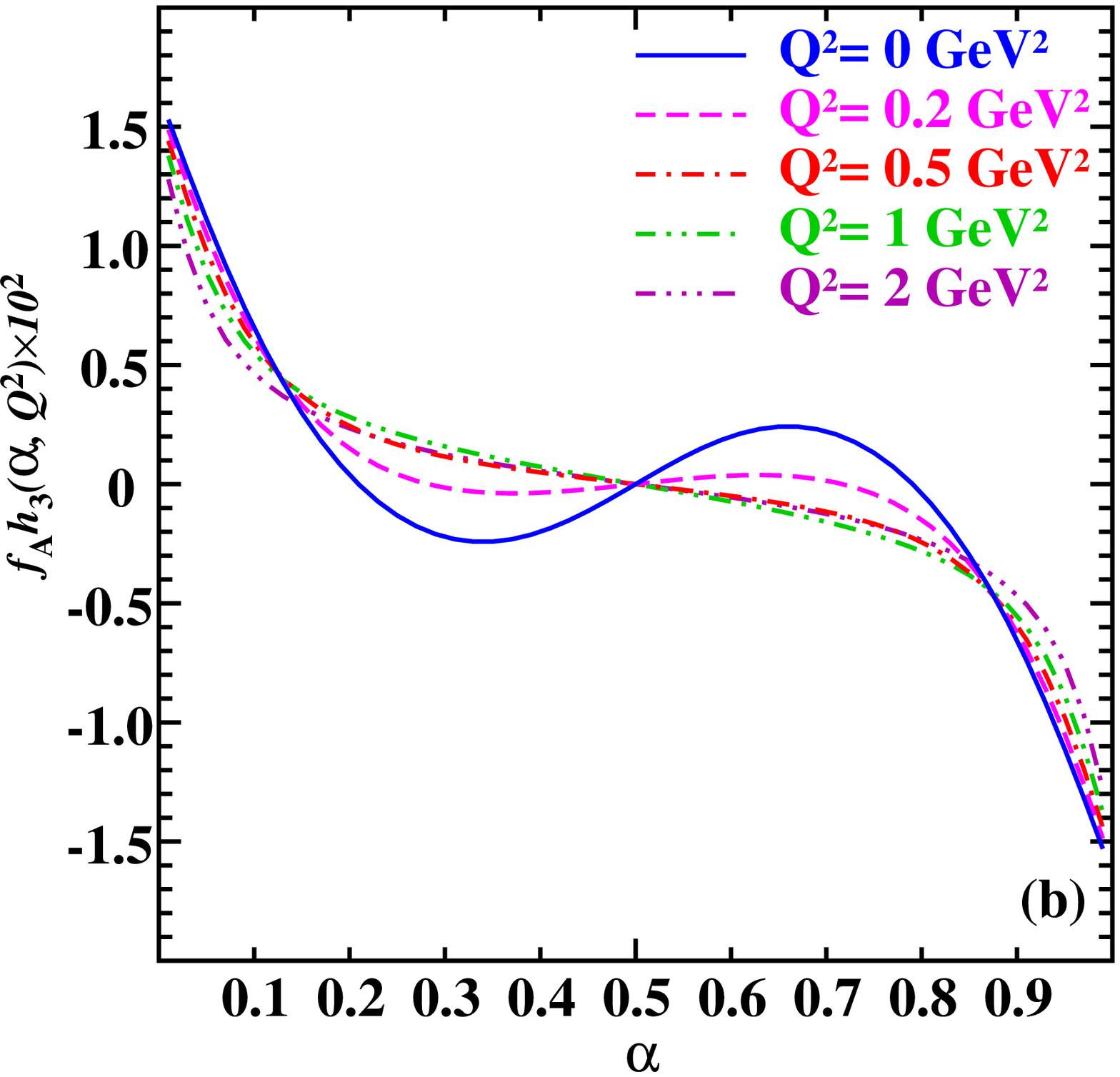}
\ec \caption{\label{fig:gh_3} (color online) (a) Dependence of the distribution
amplitude $g_{3}$ on $\alpha$ for several values of $Q^{2}$. (b)
Dependence of the distribution amplitude $h_{3}$ on $\alpha$ for
several values of $Q^{2}$. }
\end{figure}

Straightforward evaluation yields for the DAs

\begin{eqnarray}
g_{3}\left(\alpha,\, q^{2}\right) & = & \int d^{4}x\, e^{-iqx}\int\frac{dz}{2\pi}e^{i(\alpha-0.5)p\cdot z}\left\langle 0\left|\bar{\psi}\left(-\frac{z}{2}\right)\gamma_{-}\gamma_{5}\psi\left(\frac{z}{2}n\right)J_{-}^{5}(x)\right|0\right\rangle \nonumber \\
 & = & \frac{8N_{c}}{f_{A}}\int\frac{dl^{-}d^{2}l_{\perp}}{(2\pi)^{4}}\left[\frac{2(l^{-})^{2}+l^{-}q^{2}}{\left(l^{2}+\mu^{2}(l)\right)\left(\left(l+q\right)^{2}+\mu^{2}(l+q)\right)}-\frac{2(l^{-})^{2}+l^{-}q^{2}}{\left(l^{2}+m^{2}\right)\left(\left(l+q\right)^{2}+m^{2}\right)}\right.\label{eq:g3}\\
 & + & \left.\frac{\left(\mu(l)\left(l^{-}+\frac{q^{2}}{2}\right)-\mu(l+q)l^{-}\right)}{\left((l+q)^{2}-l^{2}\right)}\frac{M\left(f(l+q)-f(l)\right)^{2}\left(2l^{-}+\frac{q^{2}}{2}\right)}{\left(l^{2}+\mu^{2}(l)\right)\left(\left(l+q\right)^{2}+\mu^{2}(l+q)\right)}\right]_{l^{+}=-\alpha q^{+}}\nonumber 
\end{eqnarray}

\begin{eqnarray}
h_{3}\left(\alpha,\, q^{2}\right) & = & \frac{g_{\beta\mu}p_{\nu}-g_{\beta\nu}p_{\mu}}{4f_{A}}\int d^{4}x\, e^{-iqx}\int\frac{dz}{2\pi}e^{i(\alpha-0.5)p\cdot z}\left\langle 0\left|\bar{\psi}\left(-\frac{z}{2}n\right)\sigma_{\mu\nu}\gamma_{5}\psi\left(\frac{z}{2}n\right)J_{\beta}^{5}(x)\right|0\right\rangle \nonumber \\
 & = & \frac{8N_{c}}{f_{A}}\int\frac{dl_{-}d^{2}l_{\perp}}{(2\pi)^{3}}\left[\frac{l_{-}\mu(l+q)+\left(l_{-}+q_{-}\right)\mu(l)}{\left(l^{2}+\mu^{2}(l)\right)\left(\left(l+q\right)^{2}+\mu^{2}(l+q)\right)}-\frac{\left(2l_{-}+q_{-}\right)m}{\left(l^{2}+m^{2}\right)\left(\left(l+q\right)^{2}+m^{2}\right)}\right.\label{eq:h3}\\
 & + & \left.\frac{1}{2}\frac{l_{\mu_{\perp}}^{2}q^{2}}{(l+q)^{2}-l^{2}}\frac{M\left(f(l+q)-f(l)\right)^{2}}{\left(l^{2}+\mu^{2}(l)\right)\left(\left(l+q\right)^{2}+\mu^{2}(l+q)\right)}\right]_{l^{+}=-\alpha q^{+}}\nonumber 
\end{eqnarray}
In Figure~\ref{fig:gh_3} the distribution amplitudes~(\ref{eq:g3},\ref{eq:h3})
are shown for several values of $Q^{2}$. Near the endpoints, these
functions are finite. Straightforward integration confirms the sum
rule~(\ref{eq:CL_sum_rule}) for the $g_{3}$.

As was discussed in Section~\ref{sec:PCAC-check}, the PCAC imposes
relations between the longitudinal DAs $h_{||}^{(p)},\, h_{||}^{(t)},g_{3}$
and corresponding pion DAs. In the Figure~\ref{fig:gh_3} we show
that these relations are indeed satisfied (small deviations are due
to the finite precision of numerical calculations). 

\begin{figure}[htb]
\bc \includegraphics[scale=0.32]{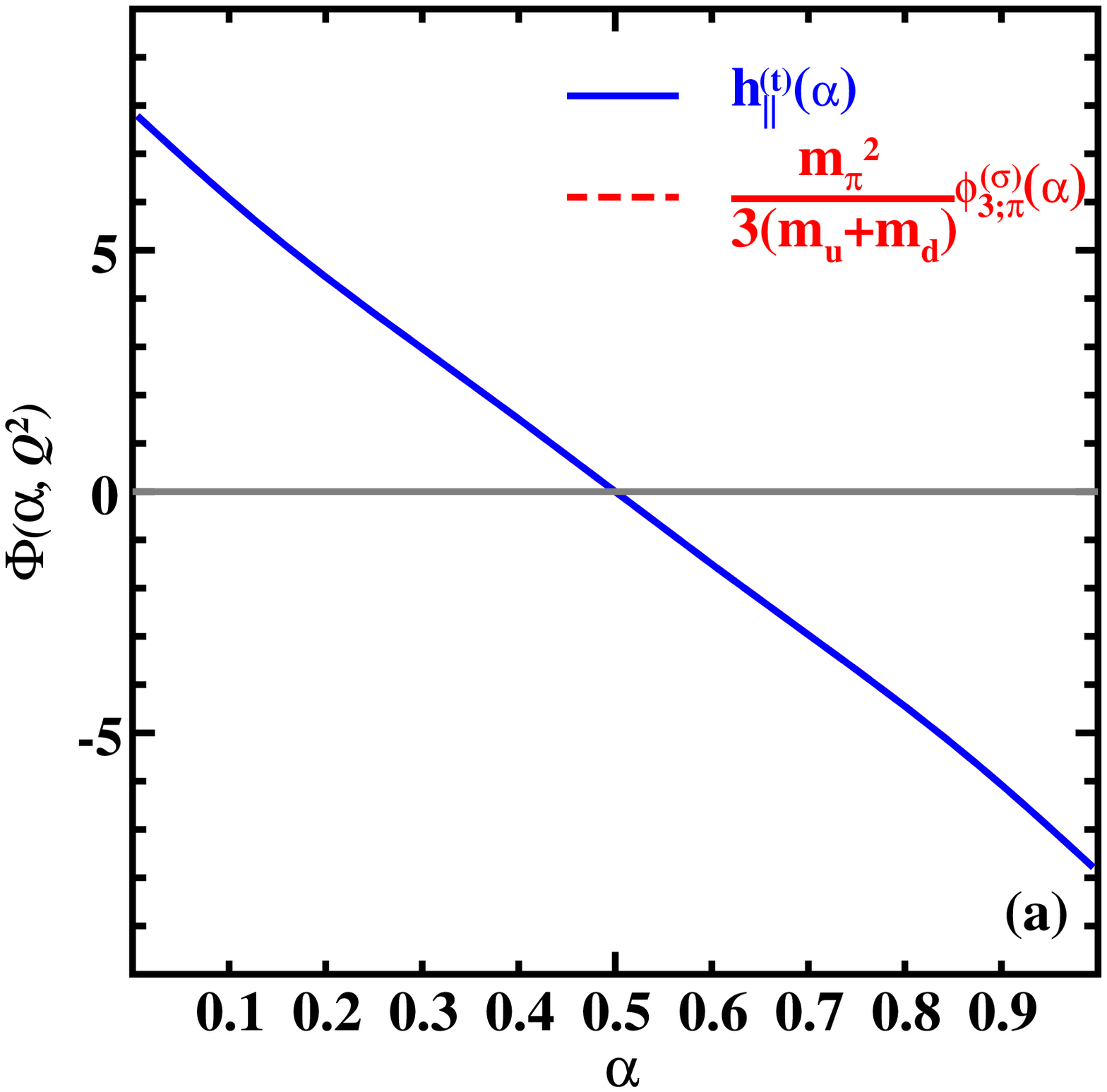}\includegraphics[scale=0.32]{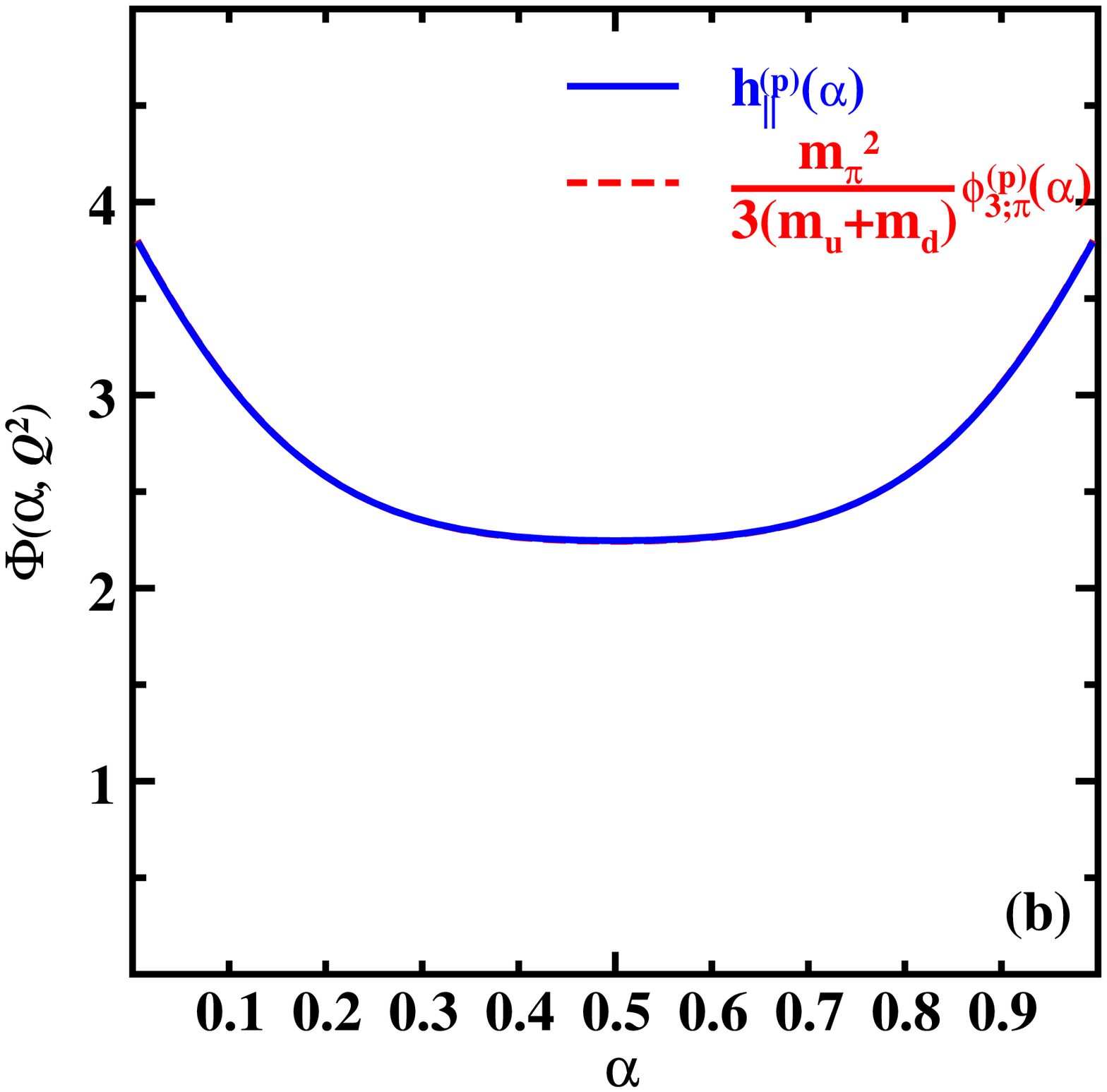}
\includegraphics[scale=0.32]{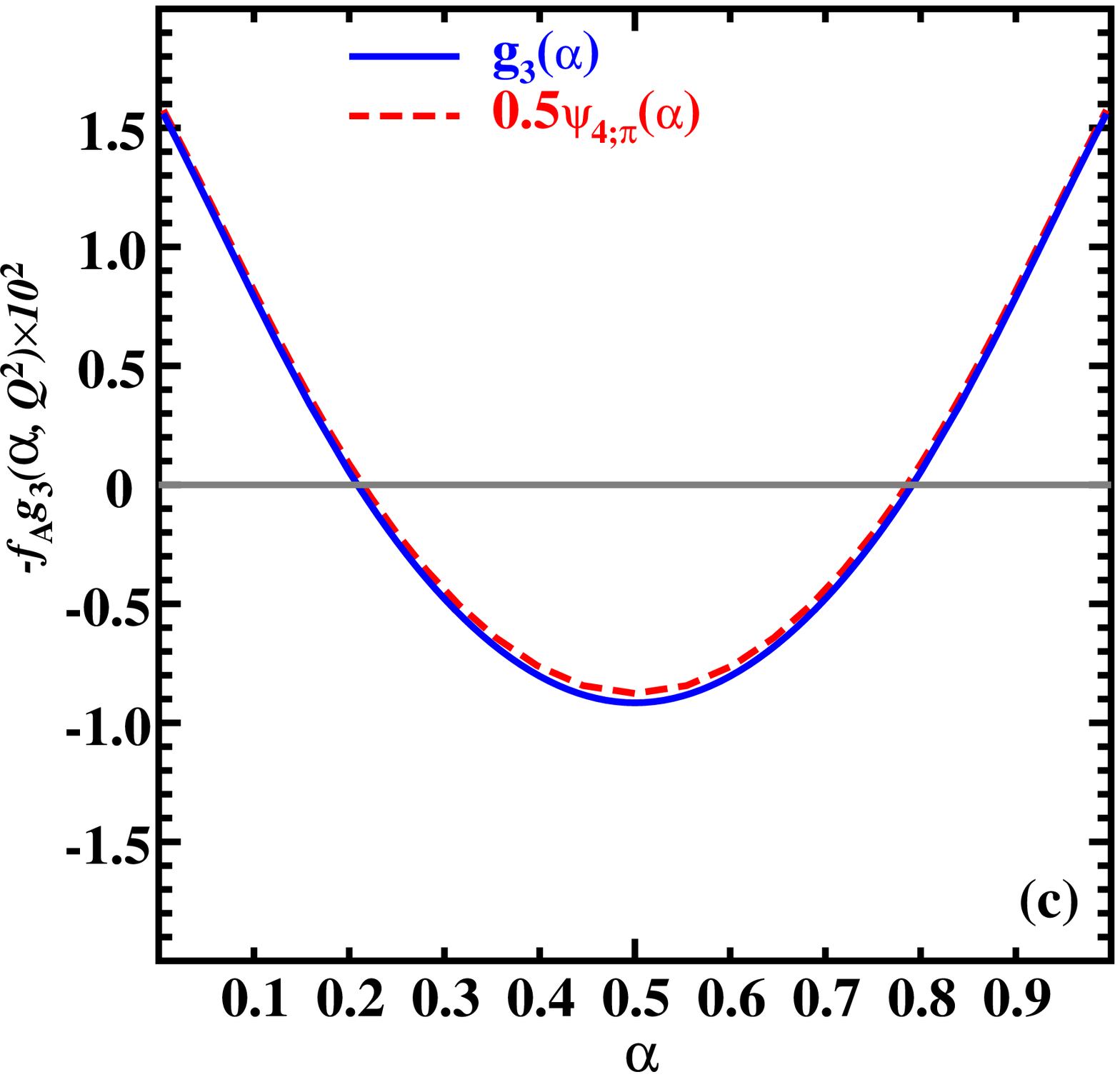} \ec \caption{\label{fig:PCAC}(color online) Comparison of the longitudinal DAs
$h_{||}^{(p)},\, h_{||}^{(t)},g_{3}$ with corresponding pion DAs.
As follows from the PCAC relations~(\ref{eq:PCAC-2}-\ref{eq:PCAC-4}),
the two curves almost coincide (deviations are less than one percent).}
\end{figure}

\section{Summary}

\label{sec:Conclusion} We studied the properties of the axial current and 
evaluated the related DAs in the
framework of the instanton vacuum model. These DAs are calculated
at the low initial scale ($\sim\rho^{-1}=600$~MeV) and can be
evolved up to a higher scale employing the standard tools. The built-in chiral
symmetry leads to the PCAC-type relations between the DAs of the
axial current and pion. 

The main result of the paper, the leading twist DAs
$\Phi_{||},\,\Phi_{\perp}$ were presented in Section~\ref{sub:Paral}.
We found that the leading-twist DA $\Phi_{||}(\alpha)$, which is
sensitive to the choice of the formfactor, has a shape close to the
asymptotic form, which vanishes at the endpoints for all possible formfactors.
The transverse DA $\Phi_{\perp}\left(\alpha\right)$ is antisymmetric
and is finite near the endpoints , as discussed in Section~\ref{sub:Phi_tr}.
For both DAs we provide simple interpolation parametrizations~(\ref{eq:Phi_Paral_interp},\ref{eq:Phi_Tr_interp}),
which allow fast numerical evaluations.

The subleading twist DAs (see Sections~\ref{sub:g_a_tr}) discussed
in Section~\ref{sub:g_a_tr}, were compared with the predictions
based on the popular Wandzura-Wilczek type relations~(\ref{eq:WW-1}-\ref{eq:WW-4}).
We found that these relations are inconsistent with prediction of
the present approach. The reason is that the twist-3 quark-gluon operators
$\bar{\psi}(x)\gamma_{\alpha}G_{\mu\nu}(x)[x,y]\psi(y)$, disregarded
in the Wandzura-Wilczek approximation, are sizable in the instanton
vacuum. Similarly, the Wandzura-Wilczek approximation fails for the
vector current DAs~\cite{Dorokhov:2006qm}. We check that the PCAC-type
relations between longitudinal DAs of the axial current and of the
pion are satisfied for all DAs.

For further practical applications we provide a computational code.

\section*{Acknowledgements}

We are grateful to A. E. Dorokhov for the fruitful discussion of the
distribution amplitudes. We acknowledge the partial support by Fondecyt
(Chile) grants No. 1090291, 1100287 and 1120920.

\appendix

\section{Pion distribution amplitudes}

\label{sec:pionDAs}

The distribution amplitudes of the pion~(\ref{eq:piWF-mu5})-(\ref{eq:piWF-munu})
and their evaluation in the framework of the IVM was performed and
discussed in detail in~\cite{Esaibegian:1989uj,Anikin:2000rq,Dorokhov:2000gu,Dorokhov:2002iu,Dorokhov:2003kf,Dorokhov:2006qm}.
For the sake of completeness we present here the leading-order expressions
for the pion distribution amplitudes.

\begin{eqnarray}
\phi_{2;\pi}\left(\alpha\right) & = & \frac{1}{if_{\pi}\sqrt{2}}\int\frac{dz}{2\pi}e^{i(\alpha-0.5)p\cdot z}\nonumber \\
 & \times & \left\langle 0\left|\bar{\psi}\left(-\frac{z}{2}n\right)\hat{n}\gamma_{5}\psi\left(\frac{z}{2}n\right)\right|\pi(q)\right\rangle =\nonumber \\
 & = & \frac{8N_{c}}{f_{\pi}\sqrt{2}}\int\frac{dl^{-}d^{2}l_{\perp}}{(2\pi)^{4}}\nonumber \\
 & \times & \left[Mf(l)f(l+q)\frac{\mu(l)\bar{\alpha}+\mu(l+q)\alpha}{\left(l^{2}+\mu^{2}(l)\right)\left(\left(l+q\right)^{2}+\mu^{2}(l+q)\right)}\right]_{l^{+}=-\alpha q^{+}}.\label{eq:phi_2;pi_0}
\end{eqnarray}

\begin{eqnarray}
\psi_{4;\pi}\left(\alpha\right) & = & \frac{\sqrt{2}}{if_{\pi}}\int\frac{dz}{2\pi}e^{i(\alpha-0.5)p\cdot z}\nonumber \\
 & \times & \left\langle 0\left|\bar{\psi}\left(-\frac{z}{2}n\right)\hat{p}\gamma_{5}\psi\left(\frac{z}{2}n\right)\right|\pi(q)\right\rangle \nonumber \\
 & = & \frac{16N_{c}}{f_{\pi}\sqrt{2}}\int\frac{dl^{-}d^{2}l_{\perp}}{(2\pi)^{4}}\nonumber \\
 & \times & \left[Mf(l)f(l+q)\frac{\mu(l)\left(l_{-}+q_{-}\right)-\mu(l+q)l_{-}}{\left(l^{2}+\mu^{2}(l)\right)\left(\left(l+q\right)^{2}+\mu^{2}(l+q)\right)}\right]_{l^{+}=-\alpha q^{+}}.\label{eq:psi_4;pi_0}
\end{eqnarray}

\begin{eqnarray}
\phi_{3;\pi}^{(p)}\left(\alpha\right) & = & \frac{1}{f_{\pi}\sqrt{2}}\frac{m_{u}+m_{d}}{m_{\pi}^{2}}\int\frac{dz}{2\pi}e^{i(\alpha-0.5)p\cdot z}\nonumber \\
 & \times & \left\langle 0\left|\bar{\psi}\left(-\frac{z}{2}n\right)\gamma_{5}\psi\left(\frac{z}{2}n\right)\right|\pi(q)\right\rangle \nonumber \\
 & = & \frac{8N_{c}}{f_{\pi}\sqrt{2}}\frac{m_{u}+m_{d}}{m_{\pi}^{2}}\int\frac{dl^{-}d^{2}l_{\perp}}{(2\pi)^{4}}Mf(l)f(l+q)\nonumber \\
 & \times & \left[\frac{\mu(l)\mu(l+q)+l^{2}+l\cdot q}{\left(l^{2}+\mu^{2}(l)\right)\left(\left(l+q\right)^{2}+\mu^{2}(l+q)\right)}\right]_{l^{+}=-\alpha q^{+}}.\label{eq:phi_3;pi_p_0}
\end{eqnarray}

\begin{eqnarray}
\phi_{3;\pi}^{(\sigma)}\left(\alpha\right) & = & \frac{3i}{2\sqrt{2}f_{\pi}}\frac{m_{u}+m_{d}}{m_{\pi}^{2}}\left(p_{\mu}n_{\nu}-p_{\nu}n_{\mu}\right)\int\frac{dz}{2\pi}e^{i(\alpha-0.5)p\cdot z}\nonumber \\
 & \times & \left\langle 0\left|\bar{\psi}\left(-\frac{z}{2}n\right)\sigma_{\mu\nu}\gamma_{5}\psi\left(\frac{z}{2}n\right)\right|\pi(q)\right\rangle \nonumber \\
 & = & -\frac{24N_{c}}{f_{\pi}\sqrt{2}}\frac{m_{u}+m_{d}}{m_{\pi}^{2}}\int\frac{dl^{-}d^{2}l_{\perp}}{(2\pi)^{4}}Mf(l)f(l+q)\nonumber \\
 & \times & \left[\frac{q_{+}l_{-}+\frac{q^{2}}{2}\alpha}{\left(l^{2}+\mu^{2}(l)\right)\left(\left(l+q\right)^{2}+\mu^{2}(l+q)\right)}\right]_{l^{+}=-\alpha q^{+}}.\label{eq:phi_3;pi_sigma_0}
\end{eqnarray}

\section{Interpolating expressions for the leading twist distribution amplitudes}

\label{sec:LTDAApprox}The distribution amplitudes $\Phi_{||},\Phi_\perp$ are given by Eqs.~(\ref{eq:Phi_parallel_0},\ref{eq:phi_tr}). 
For numerical calculations involving these DAs, it is convenient to have approximate interpolating expressions.
In this section we construct such interpolation using expansion over the Gegenbauer polynomials. 

For the DA $\Phi_{||}$ such expansion has a form
\begin{equation}
\Phi_{||}\left(\alpha,\, Q^{2},\mu^{2}\approx\rho^{-2}\right)=6x(1-x)\sum_{n=even}a_{n}\left(\mu^{2}\approx\rho^{-2},Q^{2}\right)C_{n}^{3/2}(2x-1).\label{eq:Phi_Paral_interp}
\end{equation}
Since the DA $\Phi_{||}$ in the small-$Q^2$ region is close to the asymptotic form,  we may approximate it taking the first few terms in a series~(\ref{eq:Phi_Paral_interp}). The coefficients $a_n$ in the region $Q^2\lesssim 1\,$GeV$^2$ may be approximated as
\begin{eqnarray}
a_{0}\left(\mu^{2}\approx\rho^{-2},Q^{2}\right) & \approx & \frac{0.189}{Q^{2}+0.177}-0.068,\label{eq:phi2long_a0}\\
a_{2}\left(\mu^{2}\approx\rho^{-2},Q^{2}\right) & \approx & \frac{0.071}{Q^{2}+0.665}-0.034,\label{eq:phi2long_a2}\\
a_{4}\left(\mu^{2}\approx\rho^{-2},Q^{2}\right) & \approx & \frac{0.028}{Q^{2}+0.496}-0.039,\label{eq:phi2long_a4}\\
a_{6}\left(\mu^{2}\approx\rho^{-2},Q^{2}\right) & \approx & \frac{0.21}{Q^{2}+2.5}-0.081,\label{eq:phi2long_a6}\\
a_{8}\left(\mu^{2}\approx\rho^{-2},Q^{2}\right) & \approx & \frac{0.347}{Q^{2}+3.939}-0.083,\label{eq:phi2long_a8}
\end{eqnarray}
where we have explicitly shown dependence on the intrinsic scale $\mu^{2}\sim\rho^{-2}\approx0.6$~GeV$^{2}$
to avoid confusion with virtuality $Q^{2}$. As we can see from~(\ref{eq:phi2long_a0}-\ref{eq:phi2long_a8}),
at small $Q^{2}$ the series coefficients decrease rapidly and the omitted terms give a negligible correction. 

Similarly, the Gegenbauer expansion of the DA $\Phi_{\perp}(\alpha,q^{2})$
has a form 
\begin{eqnarray}
\Phi_{\perp}\left(\alpha,\, Q^{2},\mu^{2}\approx\rho^{-2}\right) & = & 6x(1-x)\sum_{n=odd}a_{n}\left(\mu^{2}\approx\rho^{-2},Q^{2}\right)C_{n}^{3/2}(2x-1),\label{eq:phiTr_gegenbauer}
\end{eqnarray}
where due to to asymmetry ony odd polynomials contribute. However,
since the DA $\Phi_{\perp}(\alpha,q^{2})$ does not vanish at the
endpoints, the Gegenbauer series~(\ref{eq:phiTr_gegenbauer}) converges
very slowly. In order to have a simple and practical approximation,
we subtract the linear function $l(x)=c\,(2x-1)$ and end up with
an approximate parametrization
\begin{eqnarray}
\Phi_{\perp}\left(\alpha,\, Q^{2},\mu^{2}\approx\rho^{-2}\right) & \approx & c\,(2x-1)+6x(1-x)\sum_{n=odd}\tilde{a}_{n}\left(\mu^{2}\approx\rho^{-2},Q^{2}\right)C_{n}^{3/2}(2x-1)\label{eq:Phi_Tr_interp}
\end{eqnarray}

where
\begin{eqnarray}
\tilde{a}_{1}\left(\mu^{2}\approx\rho^{-2},Q^{2}\right) & \approx & 2.195-\frac{5.645}{Q^{2}+2.069},\label{eq:phi2tr_a1}\\
\tilde{a}_{3}\left(\mu^{2}\approx\rho^{-2},Q^{2}\right) & \approx & \frac{0.703}{Q^{2}+0.585}-\frac{3.629}{\left(Q^{2}+1.765\right)^{2}},\label{eq:phi2tr_a3}\\
\tilde{a}_{5}\left(\mu^{2}\approx\rho^{-2},Q^{2}\right) & \approx & \frac{0.0056}{Q^{2}+0.216}-0.038,\label{eq:phi2tr_a5}\\
c & \approx & -3.17.
\end{eqnarray}
and the omitted terms give a negligible correction. 

 \end{document}